%% file: main.tex
\documentclass[graybox]{svmult}

\usepackage{type1cm}        
\usepackage{makeidx}         
\usepackage{graphicx}        
\usepackage{multicol}        
\usepackage[bottom]{footmisc}

\usepackage{newtxtext}       %
\usepackage[varvw]{newtxmath}       

\makeindex             

\usepackage{caption}
\usepackage{tabularx,booktabs}
\usepackage{adjustbox}

\usepackage{color}
\usepackage{amsmath}
\usepackage{amsfonts}

\usepackage[T1]{fontenc}
\usepackage{textcomp}
\usepackage{psfrag}
\usepackage{multimedia}
\usepackage{arydshln}
\usepackage{fancybox}
\usepackage{float}
\usepackage{tikz}
\usepackage{pgfplots}
\pgfplotsset{compat=1.18} 
\usepackage{pgfplotstable}
\usetikzlibrary{pgfplots.groupplots}
\usetikzlibrary{pgfplots.dateplot}
\usetikzlibrary{external}

\iftrue
    \tikzsetexternalprefix{figures/}
\fi

\usepackage{acronym}
\usepackage{siunitx}
\usepackage{url}
\usepackage[capitalise]{cleveref}
\usepackage{xcolor}

\newcommand{\vect}[1]{\ensuremath{\boldsymbol{\mathrm{#1}}}}

\newcounter{lastnote}

\input{acronyms.tex}

\begin{document}

\title*{Data-Driven Domestic Flexible Demand: Observations from experiments in cold climate}
\titlerunning{Data-Driven Domestic Flexible Demand}
\author{Dirk Reinhardt, Wenqi Cai, and Sebastien Gros}
\institute{Dirk Reinhardt \at Department of Engineering Cybernetics, Norwegian University of Science and Technology, Trondheim, Norway, \email{dirk.p.reinhardt@ntnu.no}
    \and Wenqi Cai \at Department of Engineering Cybernetics, Norwegian University of Science and Technology, Trondheim, Norway, \email{wenqi.cai@ntnu.no}
    \and Sebastien Gros \at Department of Engineering Cybernetics, Norwegian University of Science and Technology, Trondheim, Norway, \email{sebastien.gros@ntnu.no}}
%
%
\maketitle

\abstract*{
    In this chapter, we report on our experience with domestic flexible electric energy demand based on a regular
    commercial (HVAC)-based heating system in a house. Our focus is on investigating the predictability of the
    energy demand of the heating system and of the thermal response when varying the heating system settings. Being
    able to form such predictions is crucial for most flexible demand algorithms. We will compare several methods
    for predicting the thermal and energy response, which either gave good results or which are currently promoted
    in the literature for controlling buildings. We will report that the stochasticity of a house response is--in
    our experience--the main difficulty in providing domestic flexible demand from heating. The experiments were
    carried out on a regular house in Norway, equipped with four air-to-air Mitsubishi heat pumps and a
    high-efficiency balanced ventilation system. The house was equipped with multiple IoT-based climate sensors,
    real-time power measurement, and the possibility to drive the HVAC system via the IoT. The house is operating on
    the spot market (Nord Pool NO3) and is exposed to a peak energy demand penalty. Over a period of three years, we
    have collected data on the house (temperatures, humidity, air quality), real-time power and hourly energy
    consumption, while applying various flexible demand algorithms responding to the local energy costs. This has
    produced large variations in the settings of the heating system and energy demand, resulting in rich data for
    investigating the house response. This chapter aims at providing important insights on providing flexible demand
    from houses in cold climates.
}

\abstract{
    In this chapter, we report on our experience with domestic flexible electric energy demand based on a regular
    commercial (HVAC)-based heating system in a house. Our focus is on investigating the predictability of the
    energy demand of the heating system and of the thermal response when varying the heating system settings. Being
    able to form such predictions is crucial for most flexible demand algorithms. We will compare several methods
    for predicting the thermal and energy response, which either gave good results or which are currently promoted
    in the literature for controlling buildings. We will report that the stochasticity of a house response is--in
    our experience--the main difficulty in providing domestic flexible demand from heating. The experiments were
    carried out on a regular house in Norway, equipped with four air-to-air Mitsubishi heat pumps and a
    high-efficiency balanced ventilation system. The house was equipped with multiple IoT-based climate sensors,
    real-time power measurement, and the possibility to drive the HVAC system via the IoT. The house is operating on
    the spot market (Nord Pool NO3) and is exposed to a peak energy demand penalty. Over a period of three years, we
    have collected data on the house (temperatures, humidity, air quality), real-time power and hourly energy
    consumption, while applying various flexible demand algorithms responding to the local energy costs. This has
    produced large variations in the settings of the heating system and energy demand, resulting in rich data for
    investigating the house response. This chapter aims at providing important insights on providing flexible demand
    from houses in cold climates.
}

\section{Introduction}

In recent years, with the increased digitalization of the power system and the emphasis on sustainable energy, the role of
domestic energy management has come to the forefront. At the core of this focus is the effective utilization and
management of \ac{HVAC} systems within domestic demand response programs. Optimal \ac{HVAC} management is critical for an economically effective domestic use of energy, especially in cold climates where heating systems account for a significant portion of the total
energy consumption. For instance, in 2017 in Norway it was estimated that water and indoor-space heating accounts for 78\% of the total energy
consumption of an average household~\cite{energifaktanorge2023}. The research community has investigated optimal
\ac{HVAC} management in large-scale buildings and commercial spaces, but the same effort has not been invested yet in average households.
However, with the rapid adoption of heat pumps (air-to-air, air-to-water, ground-to-water, geothermal) and the proliferation of affordable and ubiquitous \ac{IoT} devices, the potential
for smart \ac{HVAC} management in households has grown significantly.

\begin{figure}
    \center
    \includegraphics[width=0.95\columnwidth]{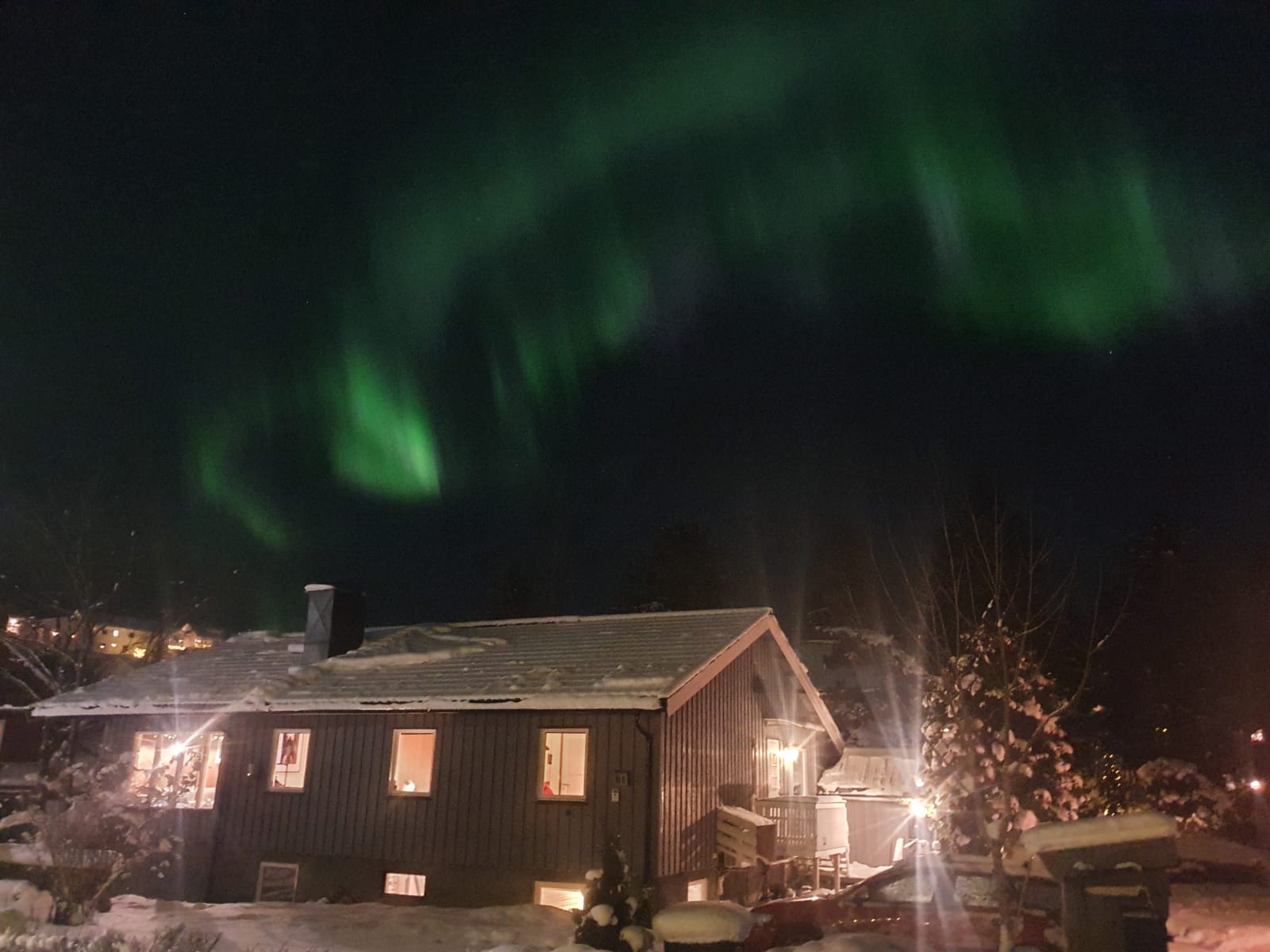}
    \caption{\footnotesize Experimental house, Jan. 2022, at latitude 63° 25' 50" N.}
    \label{fig:house}
\end{figure}

We consider the problem of domestic demand response in the context of a typical Norwegian household to be particularly
interesting. Norway has a cold climate, and the average household relies heavily on electric heating because heating
using fossil fuels (oil or gas) is prohibited, and electric heating is in general cheaper than the competing resources
(wood, pellet, biofuel). In addition, the Norwegian power system is nearly entirely hydroelectric, such that about 98\%
of the domestic electricity demand is covered by renewable energy.

Norway has created in 2001 the Nord Pool system~\cite{NordPoolGroup}, which manages the power and energy trading in a
large part of Europe. Norway and the neighbouring Scandinavian countries give access to private citizens to their local
hourly day-ahead spot prices for electric energy, allowing them to buy electricity at its real cost, and to understand
the intrinsically volatile value of electric energy. It then becomes interesting for households to manage their energy
demand according to the spot prices, by shifting it towards the lower spot prices. We will discuss the Norwegian power
system and the opportunities it offers to domestic consumers in more detail in \cref{sec:resource}. Because all
Norwegian households are equipped by law with connected smart meter that can provide real-time measurements of the power
consumption and hourly energy consumption data, it becomes fairly simple for households to monitor their demand. In
light of these developments, the Norwegian population has become increasingly conscious of its domestic energy
consumption and the related economics, and the interest in exploiting domestic flexible demand has grown significantly.
We expect this development to be a precursor of the future situation in the rest of Europe.

The day-ahead spot prices incentivize the  households to plan their energy consumption based on the future prices.
Various predictive control approaches and predictive models have been proposed to forecast the thermal response of
buildings and optimize energy consumption. This includes first-principle models based on the energy balance equations,
data-driven models based on machine learning, statistical methods, and hybrid models that combine the two
approaches~\cite{aframReviewModelingMethods2014,afroz2018modeling}. However, these models are often either difficult to
parameterize, oversimplify the complexities of the houses thermal dynamics or are difficult to use for predictive
control. Furthermore, the stochasticity of the energy response of houses poses significant challenges. In this context,
we argue in this chapter that linear data-driven models based on input-output dynamics and convex regression techniques
are best suited for implementing domestic flexible demand at large scales.

\subsection{Contribution and Outline}

In this chapter, we present an experimental study based on collected data from a typical Norwegian household depicted in
\cref{fig:house}. We first motivate the need for domestic demand response and discuss the market opportunities in
\cref{sec:resource}. We then present the experimental setup and data collection process, and discuss the challenges
associated with domestic demand response in \cref{sec:experimental_setup}. After introducing the predictive modeling
techniques used for our experiments in \cref{sec:theoretical_background}, we present their predictive performance in the
cold season of 2021-2022 in \cref{sec:results}. Our observations contribute towards establishing more efficient
demand response strategies in cold climates, underscoring the potential of informed \ac{HVAC} management in households.

\section{The role of flexible demand response in an integrated energy system}

\textcolor{black}{
    \ac{ESI} is a concept that has gained significant attention in the context of lowering the carbon footprint of
    society at large scales~\cite{omalley2016Energy}. It refers to the integration of \ac{MES}, such as electricity,
    heat, and transport, to achieve a more efficient and sustainable energy system~\cite{mancarella2014MES}. The
    integration of these systems is expected to play a crucial role in the transition to a more sustainable energy
    system, as it can help to reduce the overall energy consumption and greenhouse gas emissions. One of the key
    components of ESI is the integration of demand response, which refers to the ability of consumers to adjust their
    energy consumption in response to changes in energy prices or other signals. Considering that \ac{ESI} requires
    adaptation to the local environment and conditions~\cite{omalley2016Energy}, which may differ in economic, social,
    and environmental aspects, it is essential to understand the situation in different regions. In this chapter, we
    focus on the situation in Norway, where the power system is nearly entirely hydroelectric, and the average household
    relies heavily on electric heating. We will discuss the opportunities and challenges of domestic demand response in
    this context, before we delve into a more specific discussion with focus on integrating thermal storage and flexible
    demand of aggregated domestic consumers in the power system in the framework of \ac{ESI}.
}

\subsection{Domestic Consumption as a Resource in the Power System}
\label{sec:resource}
The power system requires that the supply of power and energy match the demand at all times and places. Complex and
expensive mechanisms are in place to ensure this balance. With the green transition, an increasingly large portion of
electric energy comes from wind and solar generation. Unlike fuel-based or hydro resources, there is no cost-effective
way to make production from wind and solar deterministic and/or adjustable on demand. Simultaneously, a growing
proportion of our energy consumption is becoming electric. Because of this shift, matching production and demand will
become more challenging in the future. Among the solutions expected to mitigate this challenge, flexible
demand is recognized as a crucial one, as long as it can be activated reliably and broadly.

In cold climates, as domestic energy demand transitions to electric energy, heating becomes the primary domestic energy consumer.
Norway, for instance, prohibited the use of fossil fuels for heating and hot water in 2018. Consequently, Norwegian
households now have one of the highest per capita electric consumption globally, with an average of over 70\% attributed to
heating. Therefore, in cold climates, flexibility in domestic demand will largely derive from heating. This flexibility
can be achieved by harnessing the thermal inertia of houses, either structurally or through heat tanks, to modulate power
and energy consumption.

To utilize flexible demand in the power system, we need efficient activation mechanisms. These mechanisms should signal
the availability of energy and power to consumers and encourage a demand that does not strain the power system
capacity. To some degree, these mechanisms already exist within various power and energy markets, rewarding those who
assist in maintaining the balance between power / energy production and demand. However, most of these mechanisms are not
currently accessible to domestic consumers due to the high minimum bidding volumes required to participate in the
markets. While this situation is anticipated to change in the future, the shift will take time.

In Scandinavia, an exception to this situation exists to some extent, given that domestic consumers can access the
day-ahead energy spot market as price-takers. Additionally, the regular transmission and distribution fees have been
updated recently to encourage domestic consumers to lower their peak hourly energy consumption. This helps in reducing
the load on distribution infrastructures. This market structure is arguably more advanced than many other regions in
Europe. It offers households the opportunity to decrease their electricity bills by leveraging this system. The exact
economic savings a household can achieve by seizing this opportunity remains uncertain and is arguably limited. The
potential savings depend on how the household is equipped, primarily in terms of its heating system, water heater,
\ac{EV} charging, and potential storage capacity. It also depends on the construction of the house, such as
insulation and thermal mass. Experts emphasize that the value derived from exploiting the day-ahead spot market is relatively
low compared to tapping into intra-day market, where uncertainties in production and consumption forecasts
are addressed. 
This contrast highlights the challenges of accurately forecasting energy
demand and scheduling energy production in our current power system and societal context.

Nevertheless, exploiting the opportunities available to domestic consumers through access to spot prices holds
significant value for society. It provides insights into domestic flexible demand and helps to identify the necessary tools to
maximize its benefits.  When creating tools to optimize domestic flexible demand against spot prices and peak
consumption penalties, it is essential to prepare the future, and ensure their adaptability to the new energy and power markets for domestic consumers that will emerge in the future.

\subsubsection{Spot market}
The experimental house used in this work operates on the local day-ahead hourly energy spot market managed by
Nord Pool~\cite{NordPoolGroup}. The Nord Pool spot market is regulating the cost of electric energy in central and northern European
countries and different regions in Norway and Sweden, see Fig. \ref{fig:NordPool}.  Every hour in a day is subject to a specific price per kWh.
Nord Pool publishes the hourly spot prices daily at around 1pm local time, providing price details for the upcoming day
that can be obtained automatically via an \ac{API}. Free and automatic data download are permitted by Nord Pool for
non-commercial exploitation. This ensures that the spot prices are known from at least 11 hours to up to 35 hours in
advance.

\begin{figure}[htbp]
    \centering
    \input{figures/spot.tex}
    \caption{\footnotesize Spot market in NO1 (black) and NO3 (red), both histograms (upper graph) and time
        series (lower graph). The experimental house is situated in NO3. }
    \label{fig:SpotMarket}
\end{figure}
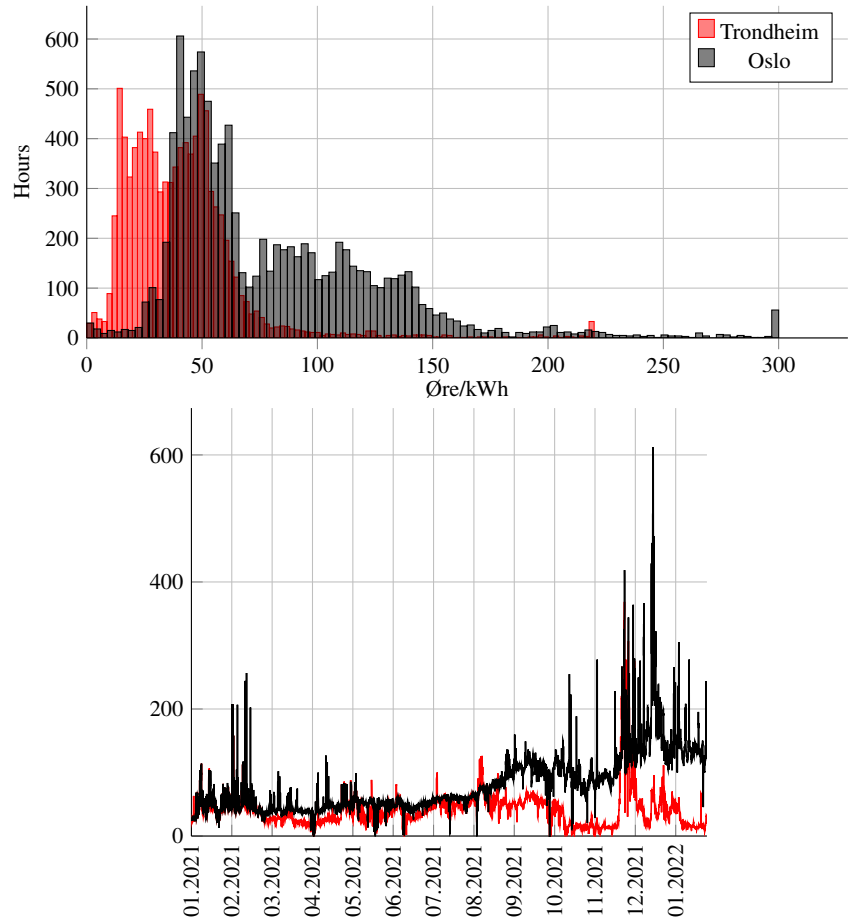

The Norwegian power system trades a substantial amount of electricity with Sweden, Germany, and the UK. Consequently,
its local spot markets are influenced by global energy prices and related contingencies. In Norway, this influence is particularly
evident in the spot markets of southern regions (price areas NO1, NO2, NO5, see \cref{fig:NordPool}). The experimental house is in the price area
NO3, which is less influenced by european prices due to the limited power transmission capacities between the north and
south Scandinavia. While the NO3 spot
market generally offers lower average prices than the rest of Europe, it does encounter frequent large price
fluctuations and occasional spikes. We illustrate this in \cref{fig:SpotMarket}.

In Scandinavia, households can access the spot market as price-takers and choose to be billed based on the local spot prices, with the
addition of a VAT and negligible administrative fees. By adjusting their consumption to avoid peak spot prices, households have the potential to lower their electricity bills. The amount of savings largely depends on the heating
system and the physical characteristics of the house. While the potential savings can justify certain investments, the extent of these investments may be limited.

\subsubsection{Grid \& Peak Demand Costs}
In addition to purchasing electric energy, households are subject to transmission and distribution costs, invoiced in
kWh, with day and night tariffs\footnote{\url{https://www.nve.no/reguleringsmyndigheten/kunde/nett/nettleie/}}. The
policy for these costs varies by region. In the area where the experimental house is located and at the time of the data
collection presented here, the policy adheres to a conventional day/night tariff, charging 36.25 NOK/kWh during the day
and 28.4 NOK/kWh at night. Furthermore, starting from July 2022, a peak power consumption fee was introduced. This fee
calculates the average of the three highest hourly energy usages within a month and bills them based on set kWh
thresholds. By imposing these peak consumption penalties, domestic users are encouraged to keep a close watch on their
top hourly demands. This indirectly helps reduce the strain on distribution infrastructures. Even if these penalties are
not substantial, they further motivate the optimization of demand management.

\begin{figure}
    \center
    \includegraphics[width=.5\textwidth,clip,trim= 0 0 0 0]{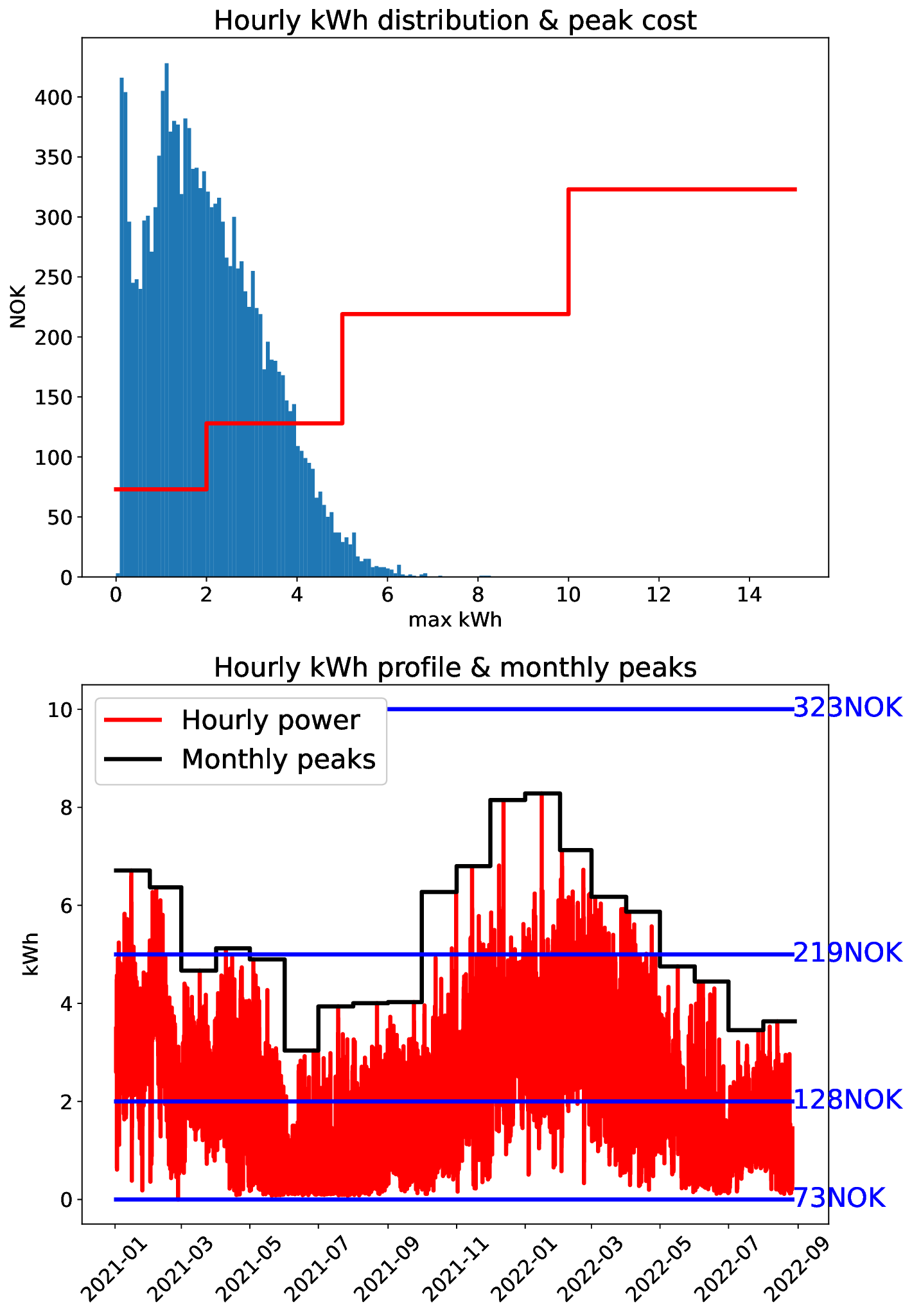} 
    \caption{ \footnotesize Monthly hourly peak energy penalty policy. The red stairs in the upper graph show the different thresholds of monthly hourly peak energy (in kWh) and the corresponding penalty (in NOK). The blue histogram shows the hourly total energy consumption in the experimental house. The lower graph shows the hourly total energy consumption (in red) and the monthly peak hourly energy (in black). The blue levels show the thresholds, numbered by their respective costs (in NOK). The current penalty policy is based on averaging the three peak energy demand in each month.}
    \label{fig:PeakPenalty}
\end{figure}

\subsubsection{Current \& Future Opportunities}

\begin{figure}
    \center
    \includegraphics[width=0.95\columnwidth]{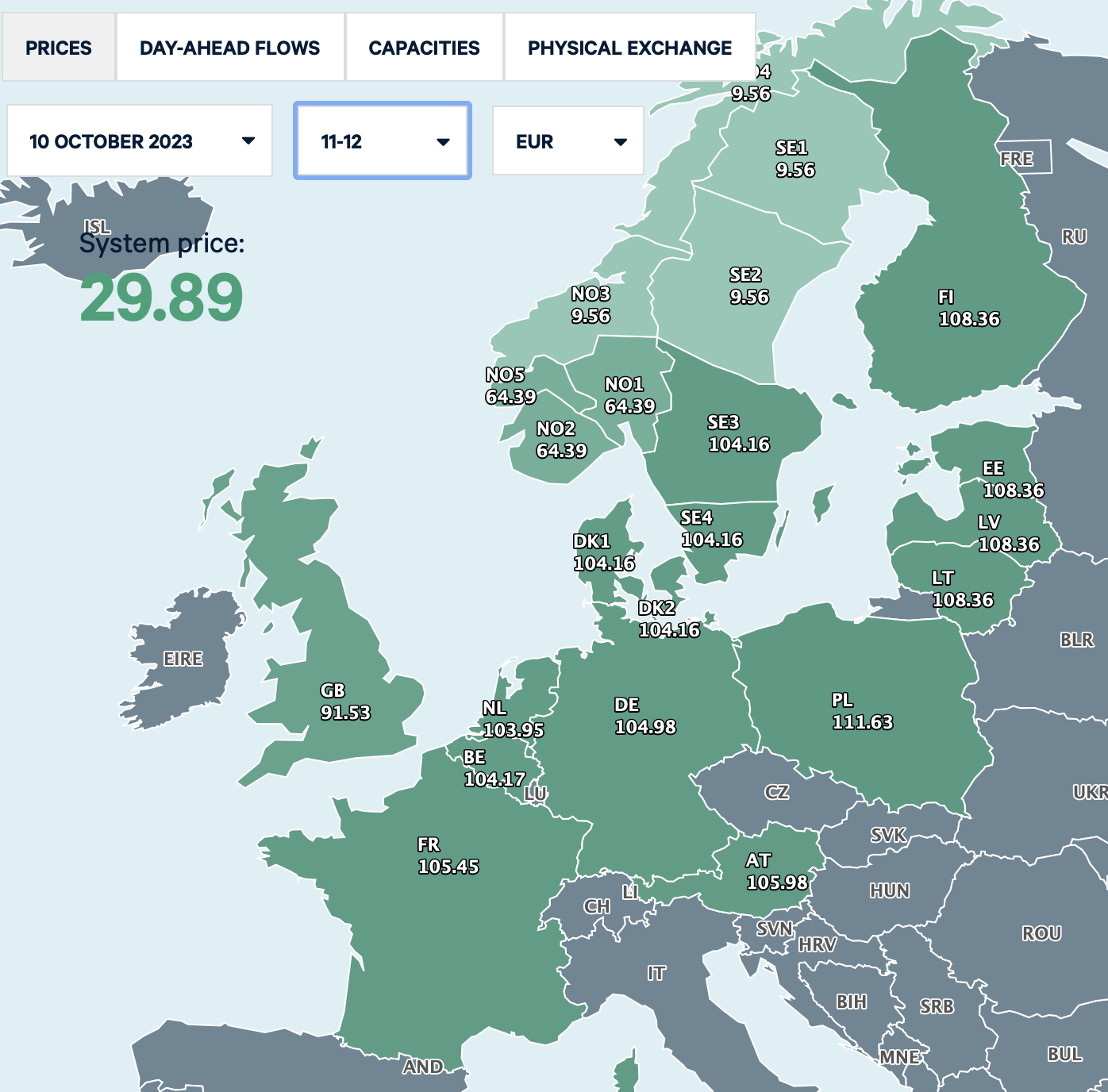}
    \caption{\footnotesize Map of the day-ahead hourly spot prices (here for the $10^\mathrm{th}$ of Oct. 2023, 11-12am)
        in the Nord Pool system. Nord Pool defines the energy and power markets for a large part of Europe (highlighted in
        green on the map).}
    \label{fig:NordPool}
\end{figure}

The current NordPool system covers a large part of Europe, see~\cref{fig:NordPool}.
Among the various power and energy markets managed by Nord Pool, day-ahead hourly spot prices are assigned for covered
areas. Whether domestic consumers can access these spot prices depends on local (national) policies. When given access
to these prices, domestic consumers have incentives to implement solutions that modify their energy demand in line with
the spot prices in their specific price area.

Various commercial products are available for managing household energy demand, commonly referred to as ``smart house"
solutions. These products are especially prevalent in Scandinavia, where electric energy consumption in households is
significant. Although the exact algorithms or optimization methods employed by these commercial solutions remain
unclear, it is reasonable to infer that many operate on straightforward principles. Some solutions focus on controlling
lights and low-power devices, while others offer user-defined thermostat scheduling for heating systems. A few might
provide rudimentary responses to spot prices, usually just slightly lowering the thermostat set point during high-price
periods. The hesitancy of commercial solutions to fully harness household energy responses is understandable.  After
all, household energy demand optimization involves delicately weighing economic benefits against comfort. Given that
different households may perceive this trade-off uniquely, crafting a one-size-fits-all solution is challenging.

Beyond the subjective nature of comfort, the implementation of domestic flexible demand, especially from heating, poses
significant challenges. These include 1) the intricacies of precisely controlling the power demand of commercial heat
pumps and 2) the challenge of accurately forecasting the thermal response of buildings and houses. A deep understanding
and management of these elements are essential for establishing predictable and dependable domestic flexible demand.
This chapter seeks to present observations and tests conducted on these challenges in a typical experimental house in
Norway.

\subsection{Domestic Flexible Demand: needs \& perspectives}

Optimally managing the flexible energy demand within a household presents some challenges, particularly when leveraging
the heating system for flexibility. Especially in the absence of substantial in-house energy storage, such as a large
heat tank or battery, this demand flexibility becomes a complex economic optimization problem, aiming at balancing
financial benefits against comfort. Comfort, as previously noted, is subjective, varying between households, individual
occupants, and even fluctuating throughout the day for a single person. As a result, tools designed for optimizing
domestic flexible demand should be able to gather feedback from occupants and integrate this feedback into their
assessment of the trade-off between economy and comfort.

Beyond varying comfort levels, we observe that the thermal response of houses and buildings is also highly stochastic
and unpredictable, \textcolor{black}{especially when relying on air-to-air heat pumps and convection in the rooms.
    Recently built houses that are well insulated and equipped with direct floor heating or geothermal heat pumps may
    exhibit less stochasticity in their thermal response. However, the majority of houses in Norway are older and rely on
    air-to-air heat pumps, and even newly built houses are often supplemented with air-to-air heat pumps due to their
    low running cost.} Simulation-based studies that explore various techniques for capitalizing on the flexible demand
of buildings often do not account for this unpredictability, leading them to overstate the efficacy of some optimization
methods. Another common misconception in numerous studies is the assumption that one can directly control the power
demand of heating systems. While this assumption may be valid for direct heating systems, which have constant power
consumption and function in a straightforward on/off manner, these systems are on the decline due to their inefficiency.
On the other hand, commercial heat pumps, such as HVAC, air-to-water, ground-to-water, and geothermal systems, do not
offer users the ability to control the power demand directly. This is due to the presence of various internal safety and
control mechanisms that ensure the system's reliability. Directly altering the power demand could sidestep these
safeguards, potentially voiding manufacturer warranties. It is unlikely that this situation will shift in the immediate
future.

The power demand of commercial heating systems can be influenced by adjusting the temperature set points given to the
unit, such as thermostat settings or heat tank temperature. However, this method of control is indirect and lacks
precision.  As a result, the actual power and energy demands are difficult to predict and should be considered
stochastic. This characteristic has been noted in multiple scenarios, especially evident for HVAC-based heating systems,
which are common in Scandinavia, as discussed in~\cref{sec:HPPower}. Effective flexible demand will require tools that
are able to handle the stochastic thermal and energy responses of houses and buildings.

Due to the small power and energy volumes associated with individual domestic consumption, realizing the potential of
domestic flexible demand in the power system necessitates large-scale deployment across households, optimal utilization
of each household's flexibility, and coordination on a large scale. To encourage widespread adoption, hardware and
software tools designed for this purpose must be cost-effective. Consequently, algorithms that facilitate domestic
flexible demand should not entail any extensive engineering or technical efforts during installation in individual
homes, as high investment costs could deter potential users. Therefore, the use of dependable data-driven methods,
capable of customizing algorithms for flexible demand to individual households, is crucial. Emphasis should be placed on
methods rooted in convex regression and optimization to ensure convergence to global minima. Additionally, these methods
should have robust statistical properties to ensure reliable outcomes.

Optimization methods designed to leverage flexible demand should be highly adaptive to the evolving power and energy
markets. As the proportion of wind and solar energy in our energy mix grows, power and energy demands shift, and the
digitalization of the power system advances. Consequently, we anticipate significant alterations in power and energy
trading methods. Current evolutions include new tariffs encouraging decreased peak energy consumption at the domestic
level, reductions in the required volumes for bidding in energy and power markets, and plans to shorten the bidding
interval from one hour to 15 minutes. Therefore, tools designed for flexible demand should operate in direct accordance
with the prevailing power and energy trading markets, ensuring agility in adapting to rapid market changes.  This
necessitates tools that can explicitly factor in economic costs and benefits. \textcolor{black}{The inertia of the thermal
    response of houses and buildings, including the heat pumps or other heating systems, is a key factor in the
    implementation of flexible demand. A relevant question in the context of reduced bidding intervals is whether the inertia
    of the heat pumps or other heating systems is sufficiently small to provide flexibility at the 15-minute scale. In our
    experience, the inertia of the heat pumps to setpoint changes is not significant, but forming predictions of the power
    consumption on this timescale can be a challenge. This is a topic that we will address in \cref{sec:HPPower}.}

\ac{EMPC} has been recognized as a
promising approach for managing complex flexible demand in many publications \cite{killianTenQuestionsConcerning2016,seraleModelPredictiveControl2018,liangMPCControlImproving2015,zhanDataRequirementsPerformance2021,oldewurtelUseModelPredictive2012,liReviewBuildingEnergy2014,mirakhorliOccupancyBehaviorBased2016}.
Meeting this need makes the use of black-box optimization methods, like \ac{DRL}, less favorable, even though it is
touted as a primary solution for energy management in buildings and houses \cite{Natale2022LessonsLF}. The reason is that \ac{DRL}
algorithms would demand complete retraining in every building and house whenever the economics of flexible demand alter,
compared to a straightforward cost function update in explicit optimization tools like \ac{EMPC}. Furthermore, managing
domestic flexible demand through transparent and \emph{explainable} algorithms is vital. Given the inherent trade-offs
between economics and comfort in domestic flexible demand, it is essential that occupants understand the flexible demand
algorithm's decisions to promote its acceptance. This requirement further disfavors black-box optimization strategies.

Lastly, since \ac{EMPC} can be readily employed as an optimization method in extensive and intricate multi-agent systems, it
will serve as an ideal foundation for coordinating domestic flexible demand across the power system. This will help in
aggregating the necessary volumes to realize the complete potential of domestic flexible demand.

\subsection{The role as a Multi-Energy System Resource in the Energy System Integration framework}

\ac{ESI} includes all types of energy sources and power consumption, for which \cite{omalley2016Energy} propose
three ``opportunity areas" to give a coarse differentiation between solutions to reach reliable, economic, and
environmental goals within the energy sector. We detail next how the content of this chapter can be related to
these opportunity areas.

\subsubsection{Streamlining the Energy System}

Streamlining the energy system emphasizes optimizing the integration of diverse energy sources and managing
demand. This involves restructuring, reorganizing, and modernizing existing systems or investing in new
infrastructure. Enhancing the flexibility of energy usage can offer system-wide advantages, potentially creating
new markets for innovative products and services. Moreover, efficient real-time locational markets can
incentivize both capacity and flexibility. The aggregated thermal inertia of residential and commercial
buildings at a large scale is considered as a significant resource for capacity and flexibility~\cite{sartoriFlexbuildFinalReport2023}. Developing
reliable, data-driven models to predict thermal responses is crucial for streamlining energy systems. More
specifically, it is crucial to use methods that are agnostic to the specific building or house, and that are
based on convex regression techniques, such that they can be rolled out at large scales without the need for
extensive engineering or technical efforts during installation. The experimental data presented herein will
support the development and validation of such models.

\subsubsection{Synergizing the Energy System}

Synergizing the energy system involves integrating various subsystems—such as electricity, heat, and
transportation—to foster a more efficient and sustainable overall system. This chapter focuses on leveraging
thermal storage and the flexible energy demand of aggregated domestic consumers within the power system. This
approach is particularly relevant in regions like Norway, where electric heating is widespread. The concept of
``virtual storage" is exploited, where flexibility in heating demand helps to balance the power system across a
broad spectrum of households. This applies moderate time scales that allow for a response to price signals and
end-user behavior. This form of storage, which relies on behavioral adjustments rather than significant capital
investment, offers an economically viable alternative to traditional storage solutions. The deployment of
low-cost IoT devices in a household setting, discussed in this chapter, facilitates the collection of data
crucial for developing demand management technologies.

\subsubsection{Empowering the End-User}

Empowerment in this context involves enabling consumers to actively participate in managing their energy
consumption.  Through minimal investment in accessible, low-cost IoT devices, average households can monitor and
control their heating demands, adapting to local energy prices optimally. This chapter's experiments aim to
provide a step towards actively using the thermal inertia of households and buildings to balance the power grid, especially in
cold climates, by predicting thermal responses and energy requirements for heating. Additionally, it explores
how data can inform users about the economic and comfort implications of their choices. Looking forward, the
development of platforms that allow users to compare their energy efficiency investments with those of their
neighbors could further engage and empower consumers. Demonstrating the feasibility of collecting the necessary
data for such comparisons at a low cost is a pivotal aspect of this discussion.

\section{Experiments}
\label{sec:experimental_setup}

In this section we describe the house used for the experiments, including the IoT system to monitor and control the house.

\subsection{Physical setup}

\begin{figure}
    \center
    \input{figures/outdoor_temperature.tex}
    \caption{\footnotesize Distribution of the outdoor temperature in the area of the experimental house. Some level of heating is used on a large part of the year.}
    \label{fig:outdoor_temperature}
\end{figure}
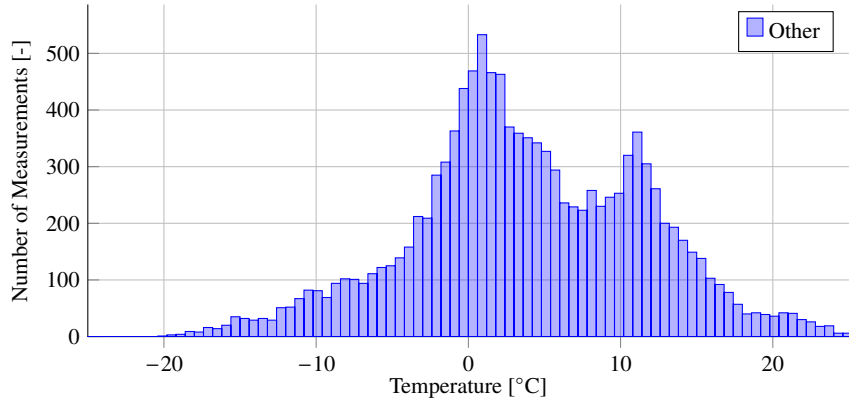

The house is located in Trondheim, which has a typically cold climate. The average yearly temperature during the
experimental period (January 2021 - March 2023) was approximately $4^\circ$C (see \cref{fig:outdoor_temperature}) and
the region typically receives significant snowfall from December to April. The house has a living space of approximately
$200\mathrm{m}^2$ and a total area of $290\mathrm{m}^2$, which includes a large workshop. The structure is made of
concrete and wood, which is typical for Norwegian households. The house is equipped with a state-of-the-art, connected
ventilation system featuring top-efficiency heat exchange. As a consequence, the windows are almost always kept closed.
The combination of the local outdoor climate and the ventilation system means that the indoor environment remains very
dry for most of the cold season. Due to the local outdoor temperatures, the heating system needs to be active or
partially active for most of the year.

The house has three full-time occupants year-round, engaging in regular activities such as cooking and taking showers.
The house also features a fireplace, which is frequently used during the winter months. Activities like cooking,
showers, and the use of the fireplace introduce unknown disturbances into the system. The hot water is supplied by an
electric hot water tank (approximately 200 liters). The water tank has a very strong insulation, resulting in minimal heat loss.

The primary heating system is based on air-to-air heat pumps. Such a setup is common in large parts of Scandinavia,
specifically Norway and Sweden, due to its efficiency, low investment cost, and ease of installation. The house includes
two outdoor heat exchangers and four indoor units (one for each volume). The heat pumps, manufactured by Mitsubishi, can
collectively achieve a peak average electric power of around 5kW. In Norway, the average \ac{COP} for air-to-air heat
pumps is estimated to be about 3. Despite its above-average size, the house used in these experiments is representative
of typical Scandinavian homes.

The house is equipped with two smart meters: one supplies power to the heating system, and the other caters to the
remaining electrical appliances. The house's power consumption can theoretically peak over 50kW. However,
this peak cannot be achieved with the currently installed appliances.

\subsection{Internet of Things (IoT)}
The house's climate and power demand are managed using an \ac{EMS} that is built around \ac{IoT} components. A key
specification for this system was its affordability at a low cost, off-the-shelf availability, and ease of installation for
average households. This design choice stems from our belief that prohibitive investment costs and a complex
installation process would significantly hinder large-scale deployment of smart-home solutions. The total cost of the
\ac{IoT} system used in the experiments discussed herein is under 500€. We next describe the details of the \ac{IoT} components.

\subsection{Power measurements}
Both smart meters supplying power to the house are equipped with devices that log real-time power demand at
$0.5\,\mathrm{Hz}$. For this purpose, we utilized the off-the-shelf device \emph{Tibber Pulse}\footnote{https://tibber.com/no/pulse}.  This
device connects to the \ac{HAN} port of the smart meters and streams power data through the house's wifi for
logging\footnote{https://www.sciencedirect.com/topics/engineering/home-area-network}.  Although the system occasionally experiences glitches, it consistently captures
sufficient data to provide a comprehensive view of the house's power consumption. Installation of the Tibber Pulse is
straightforward, allowing anyone to set it up in minutes using either a smartphone or a standard
computer\footnote{https://support.tibber.com/en/articles/4611502-pulse-han-installation}. Furthermore, the most recent hourly power consumption data (used for billing)
can be automatically retrieved from the provider whenever necessary.

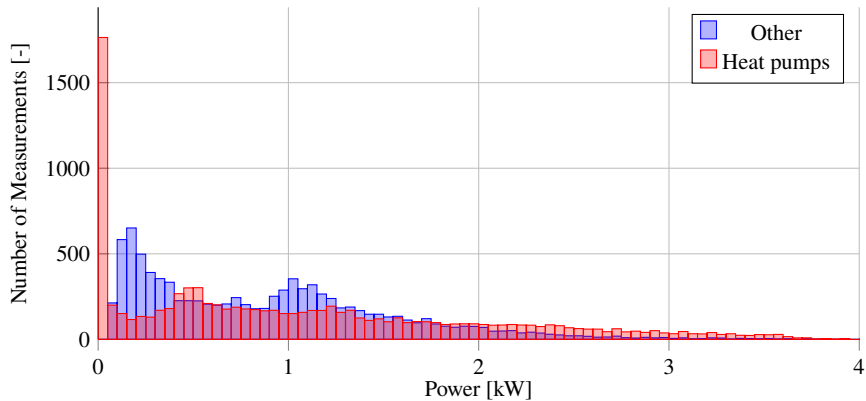
\begin{figure}
    \center
    \input{figures/power_consumption.tex}
    \caption{\footnotesize Distribution of the real-time power demand in the experimental house, heat pumps (red) and other appliances (blue).}
\end{figure}

\subsection{Heat pump control}
Mitsubishi heat pumps can be equipped with a Wi-Fi gateway and managed via MelCloud~\footnote{https://www.melcloud.com/}. However, we did not
choose this solution for the experiments due to its relatively high cost and installation complexity. Instead, a
significantly cheaper and simpler solution is to install \ac{IoT} devices that mimic the \ac{IR} remote controller of
the indoor units. These devices enable complete control over the indoor units, including functions like on/off, setting
target temperatures, and adjusting fan speed. For our experimental setup, we opted for \emph{Sensibo}
devices\footnote{https://sensibo.com/products/sensibo-sky} because of their simplicity, low cost and off-the-shelf availability.
Installing a Sensibo device is straightforward, and anyone can do it in a few minutes with a smartphone. Besides controlling
the heat pumps, these devices come equipped with climate sensors to monitor temperature and humidity. In the house,
four such devices were set up, each one managing one indoor unit and monitoring the climate in a specific ``Volume".

\subsection{Indoor climate measurements}
The \ac{EMS} primarily depends on the Sensibo devices for indoor climate measurements, specifically temperature and
humidity. However, in early 2022, we augmented the system with eight additional climate sensors and began logging data
shortly thereafter. These sensors capture temperature, humidity, and \ac{TVOC} readings. Although these measurements are
available, we regard them as supplementary and, at the time of writing, do not utilize them for decision making in the \ac{EMS}.

\subsection{Weather data}

Weather data and electricity prices are publicly available and can be obtained from various sources. We next describe
the sources of the data used in our experiments and give some insights into the market environment in which the house
operates.

The local outdoor climate data (local temperature, humidity, solar radiation) are obtained from regular weather
services.

Data for the local outdoor temperature and humidity are provided to the public by the Norwegian Meteorological Institute
(MET)~\footnote{\url{https://www.met.no/en}} and the Norwegian Centre for Climate Services~\footnote{\url{https://seklima.met.no/}}.

Solar irradiance  data are publicly available from the European Union's \ac{PVGIS}\cite{huldNewSolarRadiation2012} through
interactive web services\footnote{\url{https://re.jrc.ec.europa.eu/pvg_tools/en/}} or the provided \ac{API}. Alternative
sources of solar irradiance data are private providers such as SOLCAST~\footnote{\url{https://toolkit.solcast.com.au/}}.

\section{Theoretical background of domestic demand response modeling}
\label{sec:theoretical_background}
Modeling HVAC systems is a challenging task due to the complex interactions between the various components of the
system, the stochastic nature of the weather, behavior of the people inside the building, and possible other disturbances.
The control problem is to maintain the indoor climate at a comfortable level while minimizing the energy consumption.

Different approaches have been proposed to model HVAC systems, ranging from first-principle models to data-driven models
(black box), as discussed in~\cite{aframReviewModelingMethods2014,afroz2018modeling}. First-principle models are based
on the physical laws governing the system, and they are typically derived from the energy balance equations. The
accuracy of the physics-based models require a detailed knowledge of the system parameters, which are often difficult to
obtain.  Meaningful state-space representations of the system are also difficult to derive even for simple building
architectures. As a result the state selection is often not fully physical, and the model parameters are derived from
System Identification techniques. Because a full state measurement is not available in building and houses, estimating
the state-space model parameters becomes a non-convex problem, which can be difficult to solve reliably.

Data-driven modeling techniques that do not require explicit knowledge of the system dynamics have been proposed in the
literature. These techniques are based on the assumption that the system can be described by a \ac{LTI} model, and they
typically operate on input-output data of the system, hence circumventing the problem of not having a full state
measurement. Prominent examples are \ac{SI} methods~\cite{van2013closedloop} that have features of \ac{ARX}
models~\cite{ljungSystemIdentification1998}.  Another idea to form control-oriented predictions is to use the so-called
\ac{FL}~\cite{willems2005note} which states that the future input-output data of a system can be represented as a linear
combination of past input-output data.

This section provides a brief overview of the aforementioned modeling techniques, and it discusses their advantages and
disadvantages in the context of HVAC-based domestic demand response modeling. We further discuss the use of these
techniques in the context of predictive control strategies.

\subsection{Subspace Identification (SI)}

The method is termed ``subspace" because SI entails mapping high-dimensional data onto a
lower-dimensional subspace to distill the dynamic behavior of the system. The idea stems from predicting multiple steps
of the \ac{LTI} model in state-space form
\begin{align}
    \label{eq:LTI_sys_innovation}
     & \mathbf{x}_{k + 1} = \mathbf{A}\mathbf{x}_k + \mathbf{B}\mathbf{u}_k + \mathbf{K}\mathbf{e}_k, \\
    \label{eq:LTI_sys_innovation_output}
     & \mathbf{y}_k = \mathbf{C}\mathbf{x}_k + \mathbf{D}\mathbf{u}_k + \mathbf{e}_k
\end{align}
with state $\mathbf{x}_k \in \mathbb{R}^n$, input $\mathbf{u}_k \in \mathbb{R}^m$, output $\mathbf{y}_k \in \mathbb{R}^p$, and disturbance
$\mathbf{e}_k \in \mathbb{R}^p$. The matrices $\mathbf{A} \in \mathbb{R}^{n \times n}$, $\mathbf{B} \in \mathbb{R}^{n \times m}$, $\mathbf{C} \in
    \mathbb{R}^{p \times n}$, $\mathbf{D} \in \mathbb{R}^{p \times m}$, and $\mathbf{K} \in \mathbb{R}^{n \times p}$ are the system
matrices. We do not make any assumptions on the disturbance $\mathbf{e}_k$ for now.

The equivalent predictor form with $\tilde{\mathbf{A}} = \mathbf{A} - \mathbf{B}\mathbf{K}$ and $\tilde{\mathbf{B}} =
    \mathbf{B} - \mathbf{K}\mathbf{D}$ is
\begin{align}
    \label{eq:LTI_sys_prediction}
     & \mathbf{x}_{k + 1} = \tilde{\mathbf{A}}\mathbf{x}_k + \tilde{\mathbf{B}}\mathbf{u}_k + \mathbf{K}\mathbf{y}_k, \\
     & \mathbf{y}_k = \mathbf{C}\mathbf{x}_k + \mathbf{D}\mathbf{u}_k + \mathbf{e}_k,
\end{align}
\textcolor{black}{which is the form used in the \ac{SI} methods. It is obtained by substituting the output equation \eqref{eq:LTI_sys_innovation_output} into the state equation
    \eqref{eq:LTI_sys_innovation} via the error $\mathbf{e}_k$.}

The problem of finding data-driven predictors is to represent the future output $\hat{\mathbf{y}}_{k + N_p + 1}$ as a
function of the preceding input-output data,
\begin{equation}
    \{\mathbf{u}_{k},\dots,\mathbf{u}_{k+N_p}, \mathbf{y}_{k},\dots,\mathbf{y}_{k+N_p}\},
\end{equation}
giving a one-step linear predictor. If sequence of future inputs is available, i.e.
$\{\mathbf{u}_{k+N_p+1},\dots,\mathbf{u}_{k+N_p+N_f}\}$, multi-step predictors can be constructed, giving:
\begin{equation*}
    \{\hat{\mathbf{y}}_{k+N_p+1},\dots,\hat{\mathbf{y}}_{k+N_p+N_f}\}.
\end{equation*}
It is not necessary to find the state-space matrices $\mathbf{A}$, $\mathbf{B}$, $\mathbf{C}$, $\mathbf{D}$, and $\mathbf{K}$ to construct the
predictors. In this section, we discuss the central elements of \ac{SI} methods~\cite{katayamaSubspaceMethodsSystem2005,van2013closedloop}.

Given an output sequence, we adopt the notation in \cite{van2013closedloop} and define the Hankel matrix as
\begin{equation}
    \label{eq:hankel_matrix}
    \mathbf{Y}_{i,s,\bar{N}} = \left[ {\begin{array}{cccc}
                    \mathbf{y}_i           & \mathbf{y}_{i + 1} & \ldots & \mathbf{y}_{i + \bar{N} - 1}     \\
                    \mathbf{y}_{i + 1}     & \mathbf{y}_{i + 2} & \cdots & \mathbf{y}_{i + \bar{N}}         \\
                    \vdots                 & \vdots             & \ddots & \vdots                           \\
                    \mathbf{y}_{i + s - 1} & \mathbf{y}_{i + s} & \cdots & \mathbf{y}_{i + \bar{N} + s - 2}
                \end{array}} \right],
\end{equation}
where $i$ is the first index of the sequence that appears in the matrix, $s$ is the number of rows, and $\bar{N}$ is the
number of columns. We construct Hankel matrices based on input data, i.e. $\mathbf{U}_{i,s,{\bar{N}}}$ in the same way.

Now, consider the propagation of the system dynamics \eqref{eq:LTI_sys_prediction} from
time $k$ to $k+N_p$. Using the extended controllability matrix
\begin{equation}
    {\boldsymbol{\mathcal{K}}_{(\tilde{\mathbf{B}})}} = \left[
        {\begin{array}{llll} {{{\tilde{\mathbf{A}}}^{p - 1}}\tilde{\mathbf{B}}}&{{{\tilde{\mathbf{A}}}^{p - 2}}\tilde{\mathbf{B}}}& \ldots &{\tilde{\mathbf{B}}} \end{array}}
        \right].
\end{equation}
and $\boldsymbol{\mathcal{K}} = \left[\boldsymbol{\mathcal{K}}_{({\bar{\mathbf{B}}})}\quad \boldsymbol{\mathcal{K}}_{({\mathbf{K}})}\right] \in \mathbb{R}^{n \times (m + p)N_p}$,
we can write the state at time $k+N_p$ as
\begin{equation}
    \mathbf{x}_{k+N_p} = \tilde{\mathbf{A}}^{N_p}\mathbf{x}_k + \boldsymbol{\mathcal{K}}\left[\begin{array}{c}\mathbf{U}_{k,N_p,1}\\\mathbf{Y}_{k,N_p,1}\end{array}\right]
\end{equation}

\newcommand{\norm}[1]{\left\lVert#1\right\rVert}
A common assumption in \ac{SI} is that for a stable $\tilde{\mathbf{A}}$, we can choose a sufficiently long horizon $N_p$ such that
$\norm{\tilde{\mathbf{A}}^{N_p}}_F \approx 0$ \cite{van2013closedloop}. The output after $N_p$ steps can then be written as
\begin{equation}
    \mathbf{y}_{k+N_p} = \mathbf{C}\boldsymbol{\mathcal{K}}\left[\begin{array}{c}\mathbf{U}_{k,N_p,1}\\\mathbf{Y}_{k,N_p,1}\end{array}\right] + \mathbf{D}\mathbf{u}_{k+N_p} + \mathbf{e}_{k+N_p}.
\end{equation}

Now consider the propagation of the system dynamics \eqref{eq:LTI_sys_prediction} from time $k+N_p$ to $k+N_p+N_f$. Let
the extended observability matrix $\boldsymbol{\Gamma} \in \mathbb{R}^{pN_f \times n}$ and Toeplitz matrix $\mathbf{H}_{(\mathbf{B},\mathbf{D})} \in \mathbb{R}^{pN_f \times (m + p)N_f}$ be denoted as
\begin{equation}
    \label{eq:observability matrix}
    \boldsymbol{\Gamma} = \left[ {\begin{array}{c}
                    \mathbf{C}               \\
                    \mathbf{C}\mathbf{A}     \\
                    \mathbf{C}{\mathbf{A}^2} \\
                    \vdots                   \\
                    \mathbf{C}{\mathbf{A}^{N_f - 1}}
                \end{array}} \right].
\end{equation}
and
\begin{equation}
    \label{eq:toeplitz_matrix}
    \mathbf{H}_{(\mathbf{B},\mathbf{D})} = \left[ {\begin{array}{ccccc}
                    \mathbf{D}                                 & 0                                          & 0      & \ldots               & 0          \\
                    \mathbf{C}\mathbf{B}                       & \mathbf{D}                                 & 0      & \ldots               & 0          \\
                    \vdots                                     & \vdots                                     & \vdots & \vdots               & \vdots     \\
                    \mathbf{C}{\mathbf{A}^{N_f - 2}}\mathbf{B} & \mathbf{C}{\mathbf{A}^{N_f - 3}}\mathbf{B} & \ldots & \mathbf{C}\mathbf{B} & \mathbf{D}
                \end{array}} \right].
\end{equation}
with $\mathbf{H}_{(\mathbf{K},\mathbf{I})}\in \mathbb{R}^{pN_f \times (m + p)N_f}$ constructed in the same way as $\mathbf{H}_{(\mathbf{B},\mathbf{D})}$.

Then the output up to time $k+N_p+N_f-1$ can be written as
\begin{equation}
    \label{eq:data_equation_lite}
    \mathbf{Y}_{k_p,N_f,1} =
    \boldsymbol{\Gamma}\boldsymbol{\mathcal{K}}\left[\begin{array}{c}\mathbf{U}_{k,N_p,1}\\\mathbf{Y}_{k,N_p,1}\end{array}\right] + \mathbf{H}_{({\mathbf{B},\mathbf{D}})}\mathbf{U}_{k_p,N_f,1} + \mathbf{H}_{(\mathbf{K},\mathbf{I})}\mathbf{E}_{k_p,N_f,1},
\end{equation}
with $k_p = k + N_p$. Suppose $\bar N$ input-output sequences are available to form regression problems for \ac{SI}.
Then concatenating \cref{eq:data_equation_lite} horizontally gives the data equation that is central to
subspace identification algorithms~\cite{Wingerden2022}:
\begin{equation}
    \label{eq:data_equation}
    \mathbf{Y}_{k_p,N_f,{\bar N}} =
    \boldsymbol{\Gamma}\boldsymbol{\mathcal{K}}\left[\begin{array}{c}\mathbf{U}_{k,N_p,{\bar N}}\\\mathbf{Y}_{k,N_p,{\bar N}}\end{array}\right] + \mathbf{H}_{(\mathbf{B},\mathbf{D})}\mathbf{U}_{k_p,N_f,{\bar N}} + \mathbf{H}_{(\mathbf{K},\mathbf{I})}\mathbf{E}_{k_p,N_f,{\bar N}},
\end{equation}

The matrices $\boldsymbol{\Gamma}\boldsymbol{\mathcal{K}}$ and $\mathbf{H}_{(\mathbf{B},\mathbf{D})}$ can be found by solving \eqref{eq:data_equation} in a least-squares
sense. Specifically, stacking the data matrices into
\begin{equation}
    \label{eq:Z_bar_N}
    {\mathbf{Z}_{\bar N}} = \left[ {\begin{array}{c} {{\mathbf{U}_{k,N_p,\bar N}}} \\ {{\mathbf{Y}_{k,N_p,\bar N}}} \\ {{\mathbf{U}_{{k_p},N_f,\bar N}}} \end{array}} \right],
\end{equation}
then the least-squares problem
\begin{equation}
    \label{eq:ls_problem_phi_N_f-step}
    \mathop {\min }\limits_{\left[ {\begin{array}{ll} {\boldsymbol{\Gamma}\boldsymbol{\mathcal{K}}}&{{\mathbf{H}_{(\mathbf{B},\mathbf{D})}}} \end{array}} \right]} {\left\| {{\mathbf{Y}_{{k_p},f,\bar N}} - \left[ {\begin{array}{ll} {\boldsymbol{\Gamma} \boldsymbol{\mathcal{K}}}&{{\mathbf{H}_{(\mathbf{B},\mathbf{D})}}} \end{array}} \right]{\mathbf{Z}_{\bar N}}} \right\|_F^2}{\text{,}}
\end{equation}
has the explicit solution
\begin{equation}
    \label{eq:phi_N_f-step}
    \boldsymbol{\Phi}_{N_f-{\text{step}}} := \left[ {\begin{array}{ll} {\widehat {\boldsymbol{\Gamma} \boldsymbol{\mathcal{K}}}}&{{{\hat{\mathbf{H}}}_{(\mathbf{B},\mathbf{D})}}} \end{array}} \right] = {\mathbf{Y}_{{k_p},N_f,\bar N}}\mathbf{Z}_{\bar N}^T{\left( {{\mathbf{Z}_{\bar N}}\mathbf{Z}_{\bar N}^T} \right)^{ - 1}}.
\end{equation}
with $\boldsymbol{\Phi}_{N_f-{\text{step}}} \in \mathbb{R}^{pN_f \times (m+p)N_p+mN_f}$.
Under the assumption of a sufficiently rich input signal, state-of-the-art \ac{SI} methods find efficient solutions to
this problem via QR decompositions or other factorization are discussed in the
literature~\cite{Favoreel1999,katayamaSubspaceMethodsSystem2005} which can be used to further analyse sensitivities of
the error estimates~\cite{van2013closedloop} or to reduce the effect of noise via singular value thresholding~\cite{katayamaSubspaceMethodsSystem2005}.



One issue with the solution above is that the estimates $\widehat {\boldsymbol{\Gamma} \boldsymbol{\mathcal{K}}}$ and
$\hat{\mathbf{H}}_{(\mathbf{B},\mathbf{D})}$ in general do not have the structure of the system matrices
$\boldsymbol{\Gamma}\boldsymbol{\mathcal{K}}$ and $\mathbf{H}_{(\mathbf{B},\mathbf{D})}$. The most problematic is that
$\hat{\mathbf{H}}_{(\mathbf{B},\mathbf{D})}$ is not block-Toeplitz, which means the estimated predictor is not causal such
that future inputs appear in the outputs before they are actually applied. The self-similarity of the diagonal blocks of
$\hat{\mathbf{H}}_{(\mathbf{B},\mathbf{D})}$ and the decay along the block rows of
$\hat{\mathbf{H}}_{(\mathbf{B},\mathbf{D})}$ and $\widehat{\boldsymbol{\Gamma} \boldsymbol{\mathcal{K}}}$ can be
enforced by using regularization techniques and a gradient descent approach. This is part of ongoing work and the results
will be presented in a future publication. However, we discuss initializations of the gradient descent algorithm in the
next section.

\subsection{Linear one-step predictor via SI}
\label{sec:onestep}

The one-step predictor obtained by \ac{OSSI} is by design causal and we discuss it as a special case of the multi-step predictor. Assuming that
there is no direct feedthrough from the input to the output, i.e. $\mathbf{D} = \mathbf{0}$, the one-step predictor
is given by
\begin{equation}
    \label{eq:phi_one_step}
    \boldsymbol{\Phi}_{1-{\text{step}}} := {\widehat {\boldsymbol{\Gamma} \boldsymbol{\mathcal{K}}}} = {\mathbf{Y}_{{k_p},1,\bar N}}\mathbf{Z}_{\bar N}^T{\left( {{\mathbf{Z}_{\bar N}}\mathbf{Z}_{\bar N}^T} \right)^{ - 1}},\quad\mathrm{with}\quad
    {\mathbf{Z}_{\bar N}} = \left[ {\begin{array}{c} {{\mathbf{U}_{k,N_p,\bar N}}} \\ {{\mathbf{Y}_{k,N_p,\bar N}}} \end{array}} \right].
\end{equation}


Suppose that sufficient past data is available at time $t$ to form the initial conditions based on $N_p$ samples of past
input out data. Then the one-step predictor can be used to predict the output at $t+1$ as
\begin{equation}
    \hat{\mathbf{y}}_{t+1} = \boldsymbol{\Phi}_{1-{\text{step}}}\left[\begin{array}{c}\mathbf{u}_{p}\\\mathbf{y}_{p}\end{array}\right],
\end{equation}
where we use the notation $\mathbf{u}_p$ and $\mathbf{y}_p$ defined as
\begin{align}
    \label{eq:up_yp}
    \mathbf{u}_p := \mathbf{U}_{t-N_p+1,N_p,1} = \begin{bmatrix}
                                                     {\mathbf{u}}_{t - N_p + 1} \\
                                                     {\mathbf{u}}_{t - N_p + 2} \\
                                                     \vdots                     \\
                                                     {\mathbf{u}}_{t}           \\
                                                 \end{bmatrix},\quad
    \mathbf{y}_p := \mathbf{Y}_{t-N_p+1,N_p,1} = \begin{bmatrix}
                                                     {\mathbf{y}}_{t - N_p + 1} \\
                                                     {\mathbf{y}}_{t - N_p + 2} \\
                                                     \vdots                     \\
                                                     {\mathbf{y}}_{t}           \\
                                                 \end{bmatrix}.
\end{align}

One of the advantages of the \ac{OSSI} approach is its ability to function without needing an explicit state
representation. This feature makes it ideal for complex HVAC system modeling, as \ac{OSSI} operates on data vectors rather
than on the full-state space, which is often of much higher dimensions or even not observable. Besides, the optimal
value of $\boldsymbol{\Phi}_{1-{\text{step}}}$ can be efficiently computed using \eqref{eq:phi_one_step}. This efficiency is
maintained irrespective of the size of the dataset, as the linear computation complexity remains unaffected by the
dimensions of the dataset. This feature empowers \ac{OSSI} to proficiently manage large datasets. The \ac{OSSI} method is
recognized as a standard convex problem. Uniquely, it comes with the assurance that $\mathbb{E}[\vect y-\hat{\vect
            y}]=0$. This assertion implies that the average of the prediction error is zero, thereby providing an unbiased
estimation. Such a guarantee is particularly coveted in many HVAC modeling problems, as it bolsters the long-term
validity and accuracy of the predictions. However, while \ac{OSSI} theoretically satisfies the condition $\mathbb{E}[\vect
        y-\hat{\vect y}]=0$, its one-step prediction model does not necessarily extend to reliable multi-step prediction in DPC
which demands finite multi-step prediction. This means the expected value of the deviation between a finite multi-step
simulated prediction sequence (denoted as $\hat{\vect y}^{sp}$) and the actual output is not necessarily zero, i.e.,
$\mathbb{E}[\vect y-\hat{\vect y}^{sp}]\neq0$. This discrepancy could result in suboptimal solutions in DPC.

\subsection{Linear Multi-Step Predictor via SI}
\label{sec:multistep}

The multi-step predictor obtained by \ac{SI} gives predictions for a finite horizon $N_f$ subject to a given trajectory of inputs
\begin{equation}
    \hat{\mathbf{y}}_f =
    \boldsymbol{\Phi}_{N_f-{\text{step}}}
    \left[
        \begin{array}{c}
            \vect{u}_p \\
            \vect{y}_p \\
            \vect{u}_f \\
        \end{array} \right],
    \label{eq:SI matrix form 3}
\end{equation}
with
\begin{align}
    \label{eq:uf_yf}
    \mathbf{u}_f := \mathbf{U}_{t+1,N_f+1,1} = \begin{bmatrix}
                                                   {\mathbf{u}}_{t+ 1} \\
                                                   {\mathbf{u}}_{t+ 2} \\
                                                   \vdots              \\
                                                   {\mathbf{u}}_{t + N_f}
                                               \end{bmatrix}^\top,\quad
    \mathbf{y}_f := \mathbf{Y}_{t+1,N_f+1,1} = \begin{bmatrix}
                                                   {\mathbf{y}}_{t + 1} \\
                                                   {\mathbf{y}}_{t + 2} \\
                                                   \vdots               \\
                                                   {\mathbf{y}}_{t + N_f}
                                               \end{bmatrix}.
\end{align}

We further use the short notation $\boldsymbol{\Phi}_p := \widehat{\boldsymbol{\Gamma} \boldsymbol{\mathcal{K}}}$ and $\boldsymbol{\Phi}_f :=
    \hat{\mathbf{H}}_{(\mathbf{B},\mathbf{D})}$ corresponding to the past and future data of $\boldsymbol{\Phi}_{N_f-{\text{step}}}$ in
\eqref{eq:SI matrix form 3}. $\boldsymbol{\Phi}_p$ establishes the connection between the past inputs and outputs
($\mathbf{u}_p$ and $ \mathbf{y}_p$) and the future outputs $ \hat{\mathbf{y}}_f$. It serves as a translation matrix that
morphs the historical data into a representation akin to an initial state for future output prediction. $\boldsymbol{\Phi}_f$ is a
lower block triangular matrix that associates the future inputs $\mathbf{u}_f$ with the future outputs $
    \hat{\mathbf{y}}_f$. Its lower block triangular structure signifies that each future output only depends on the present
and past inputs, aligning with the causality principle in time-series prediction.

As mentioned in the previous section, the matrix $\boldsymbol{\Phi}_f$ is not necessarily block-Toeplitz if we compute
it using the explicit solution~\eqref{eq:phi_N_f-step} of the least-squares problem \eqref{eq:ls_problem_phi_N_f-step}. One way
to obtain a block-Toeplitz matrix is to split the least-squares problem into subproblems for each block row of
\eqref{eq:data_equation} and then solve them parallelly. Using the index notation in \eqref{eq:data_equation}, we can
write the least-squares solution for block row $i$ as
\begin{equation}
    \label{eq:ls_solution_phi_N_f-step_block_row}
    \mathbf{Y}_{k+N_p+i,1{\bar N}}\mathbf{Z}_{\bar N}^T\left( {{\mathbf{Z}_{\bar N}}\mathbf{Z}_{\bar N}^T}\right)^{-1} \quad\mathrm{with}\quad
    {\mathbf{Z}_{\bar N}} = \left[ {\begin{array}{c} {{\mathbf{U}_{k,N_p,\bar N}}} \\ {{\mathbf{Y}_{k,N_p,\bar N}}} \\ {{\mathbf{U}_{{k_p},i+1,\bar N}}} \end{array}} \right],
\end{equation}
to build the non-zero elements of $\boldsymbol{\Phi}_f$ by solving \eqref{eq:ls_solution_phi_N_f-step_block_row} to obtain
\begin{align}
    i = 0:    & \left[\widehat{\mathbf{C}\mathbf{A}^i\boldsymbol{\mathcal{K}}}\right]\nonumber                                                                                                                                                                                  \\
    i = 1:    & \left[\widehat{\mathbf{C}\mathbf{A}^i\boldsymbol{\mathcal{K}}} \quad \hat{\mathbf{D}}\right]\nonumber                                                                                                                                                           \\  \nonumber                                                                                                                                                          \
    i = 2:    & \left[\widehat{\mathbf{C}\mathbf{A}^i\boldsymbol{\mathcal{K}}} \quad \widehat{\mathbf{C}\mathbf{A}\mathbf{B}} \quad \widehat{\mathbf{C}\mathbf{B}} \quad \hat{\mathbf{D}}\right] \nonumber                                                                      \\
    i \geq 3: & \left[\widehat{\mathbf{C}\mathbf{A}^i\boldsymbol{\mathcal{K}}} \quad \widehat{\mathbf{C}\mathbf{A}^{i-1}\mathbf{B}} \quad \widehat{\mathbf{C}\mathbf{A}^{i-2}\mathbf{B}} \quad\dots \quad \widehat{\mathbf{C}\mathbf{B}} \quad \hat{\mathbf{D}}\right]\nonumber
\end{align}


The SI method, like the OSSI method, conveniently eschews the necessity for an explicit state-space representation
inherent in OSSI, relying solely on historical data for system identification. A significant virtue of SI resides in its
capacity for multi-step prediction, ensuring that $\mathbb{E}[\vect y-\hat{\vect y}^{sp}]=0$. This aspect is
particularly beneficial for DPC as the performance of a DPC policy is contingent on the entire prediction sequence
rather than individual prediction points. Although SI might not rival OSSI in terms of single-step prediction accuracy,
it can deliver superior performance in terms of the average prediction level when evaluated across a multi-step range.
Moreover, the multi-step prediction mechanism inherent in SI safeguards against the accumulation of prediction errors,
even in the presence of highly noisy and stochastic data from HVAC systems. This ensures stable prediction models,
avoiding the instability encountered in single-step prediction methods. In essence, SI harmoniously integrates the
advantages of the FL method and the OSSI method. However, SI sacrifices the data-size-independent computational
efficiency intrinsic to OSSI. SI might be less efficient when grappling with large datasets compared to OSSI, yet it
retains its utility as an effective method for handling large data sets.

\subsection{Fundamental Lemma (FL)}

Another class of data-oriented approaches is rooted in the Fundamental Lemma \cite{willems2005note, kerz2023data} which
states that any input-state trajectory of a linear system can be written as a linear combination of past trajectories,
possibly in the form of a Hankel matrix. It forms the bases for data-driven simulations and control
methods~\cite{Markovsky2008} and has found renewed interest by the community in the context of data-driven control after
the work~\cite{Coulson2019}. We review the data-driven prediction method based on the Fundamental Lemma in this section.

Using the available data matrix $\mathbf{Z}_{\bar N}$ in \cref{eq:Z_bar_N}, the idea is to find a linear combination of
past input and output trajectories that matches the recent data $\mathbf{u}_p,\,\mathbf{y}_p$ and the future input
trajectory $\mathbf{u}_f$ as
\begin{align}
    \label{eq:FL_linear_combination}
    \left[ {\begin{array}{c}
                            {{\mathbf{U}_{k,N_p,\bar N}}} \\
                            {{\mathbf{Y}_{k,N_p,\bar N}}} \\
                            {{\mathbf{U}_{{k_p},N_f,\bar N}}}
                        \end{array}} \right]\mathbf{g} = \left[ {\begin{array}{c} \mathbf{u}_p \\ \mathbf{y}_p \\ \mathbf{u}_f \end{array}} \right].
\end{align}
Note that the data on left hand side of the equation is the $\mathbf{Z}_{\bar N}$ matrix in \cref{eq:Z_bar_N}. Under the
condition that the input sequence $\mathbf{u}_p$ is persistently exciting of order $N_p$ and the data is noise-free, the linear combination vector
$\mathbf{g} \in \mathbb{R}^{(m+p)N_p+mN_f}$ can found by solving \cref{eq:FL_linear_combination}. The vector $\mathbf{g}$
can then be used to predict the future output $\mathbf{y}_f$ as
\begin{equation}
    {\mathbf{Y}_{{k_p},N_f,\bar N}} \mathbf{g}         = \mathbf{y}_f.
\end{equation}

In practice, the data is noisy and \eqref{eq:FL_linear_combination} cannot be solved exactly. Regularization of $\mathbf{g}$ is
required to obtain a unique solution. This can be done by solving the following optimization problem
\begin{subequations}
    \label{eq:FL_opt}
    \begin{align}
        {\mathop {\min }\limits_{\mathbf{g}, \hat{\mathbf{y}}_f} } & \quad {\Pi(\mathbf{g})} \\ {{\text{subject to:}}}&\left[ {\begin{array}{c}
                        {{\mathbf{U}_{k,N_p,\bar N}}} \\ {{\mathbf{Y}_{k,N_p,\bar N}}} \\ {{\mathbf{U}_{{k_p},N_f,\bar N}}} \\ {{\mathbf{Y}_{{k_p},N_f,\bar N}}}\end{array}} \right]\mathbf{g}
        = \left[ {\begin{array}{c} \mathbf{u}_p \\ \mathbf{y}_p \\ \mathbf{u}_f \\ \hat{\mathbf{y}}_f \end{array}} \right].
    \end{align}
\end{subequations}
where $\Pi(\mathbf{g})$ is a regularization term that penalizes some norm of $\mathbf{g}$, often $\ell_1, \ell_2$ or
hybrid formulations, possibly after some form of projection~\cite{Dorfler2022}. A critical analysis of the
regularization approach can be found in~\cite{Mattsson2023} and the equivalence to \ac{SI} when using it with
instrumentation variables for control is discussed in~\cite{Wingerden2022}.

The Fundamental Lemma approach brings forth several notable advantages. Its foremost advantage is its
straightforwardness in terms of application. Unlike system identification methods that find state-space formulations,
the Fundamental Lemma method operates directly on observed data, eliminating the need for explicit state space
representation. Moreover, FL incorporates temporal data, using both history and future sequences, which provides a more
comprehensive representation of the system dynamics. Nonetheless, a significant drawback is that FL approach lacks
robustness in handling noisy data and stochastic system behavior. This deficiency might compromise the accuracy and
reliability of the control policies derived from it, particularly in HVAC systems where significant noise or stochastic
components exist. Another disadvantage is that at each step, the FL method needs to solve an optimization problem
\eqref{eq:FL_opt} to predict the output $\mathbf{y}_f$, which is computationally inefficient and therefore not suitable
for problems with large datasets. Thus, while the FL offers a simplified and temporally inclusive approach, careful
consideration of these limitations is crucial for effective implementation. The FL method is also not guaranteed to be
causal, and does not necessarily respect the semigroup property of the system dynamics in the sense that a sequence of
single steps does not yield the same result as the corresponding multistep. The most problematic issue is
however when it is used for predictive control and the optimization problem is mixes the regularization term and the
performance criterion of the control problem. The literature indicates that an optimal trade-off between the two is very
difficult to find and extensive tuning of the weighting between the two objectives is required when using it for HVAC as
reported in~\cite{Natale2022LessonsLF}.


\section{Heat-pump power consumption}

\label{sec:HPPower}
For exploiting a household or building flexible demand, it is crucial to be able to predict and control the energy consumption of commercial heat pumps \ac{HP}, in order to seek the trade-off between the economics of the energy consumption and the comfort of the occupants. Because the power and energy demands of commercial heat pumps cannot be controlled directly, it has to be done via the various \ac{HP} settings used to manage the indoor climate (thermostat value, heat tank reference temperature, etc.). \ac{HP} power control through its settings requires a model on how the settings impact the \ac{HP} power demand. Indeed, while the internal \ac{HP} software has control over this power, this software is not public, nor dedicated to achieve a specific or constant power for some given settings.

\begin{figure}[h]
    \center
    \includegraphics[width=1\columnwidth]{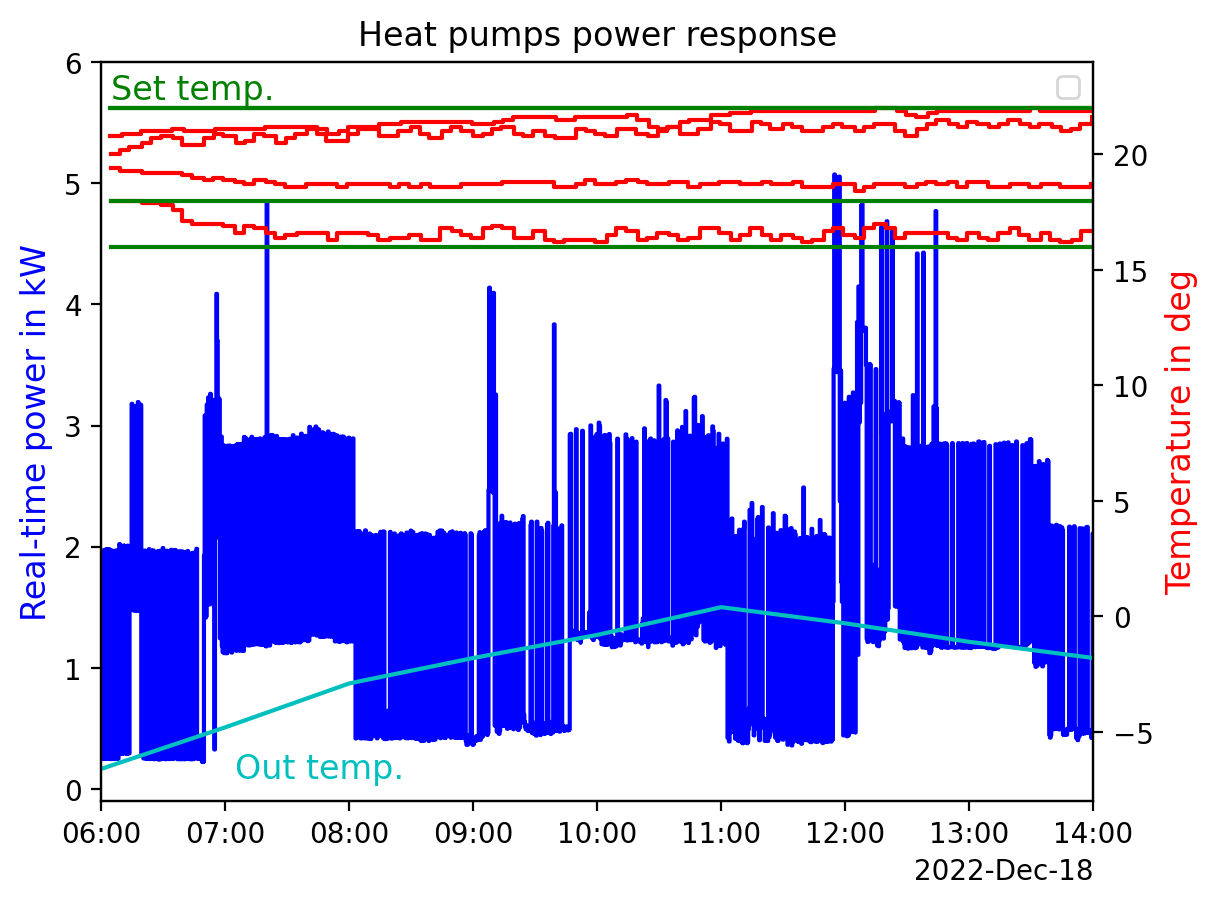}
    \caption{\footnotesize Illustration of the real-time power consumption of the heat pumps of the experimental house over an 8h period. The blue signal is the total real-time power of the heat pumps, in kW, and sampled every 2s. The red signals are the measured temperatures in the four volumes. The green lines are the temperature settings, which were constant in this time period. The fan speeds were also constant in this time period. The light blue curve displays the outdoor temperature. One can readily observe that even though the heat pumps operate in a fairly constant environment, the real-time power consumption varies a lot. Besides, the hourly consumption is not constant and does not appear correlated to the outdoor temperature variation (in light blue). This makes the real-time power consumption of the house hearing system unpredictable, while the hourly consumption is not accurately predictable.}
    \label{fig:HeatPumpPower}
\end{figure}

For heating based on HVAC (air-to-air) \ac{HP}, modelling the \ac{HP} power and energy demand as a function of the \ac{HP} settings is a challenging task. This observation is illustrated in Fig. \ref{fig:HeatPumpPower}, showing the real-time power consumption of the 4 HVAC heat pumps of the experimental house over an 8h period, for constant settings, fairly constant indoor temperatures, and a slowly varying outdoor temperature. Despite the fairly constant environment and settings, the real-time power consumption (in blue) of the \ac{HP} varies wildly. The energy consumption over different periods of time is more consistent, but it varies nonetheless significantly. These observations are not limited to this specific time window. They apply to our entire data set, regardless of the season or weather conditions. This indicates that predictions of the energy consumption of commercial HVAC (air-to-air) \ac{HP} ought to be treated as stochastic, and managed within the flexible demand optimization. It is an open question whether the energy consumption of other types of \ac{HP} (air-to-water, ground-to-water, geothermal) can be predicted more accurately.

We briefly present hereafter our observations on the modelling of the \ac{HP} hourly energy demand as a function of the
\ac{HP} settings. The choice of focusing on hourly energy demand stems from the current spot market structure, using
hourly spot prices. If the Nord Pool spot market reduces its bidding intervals to 15 minutes, as is currently being
discussed, then the \ac{HP} energy demand predictions will need to be performed also on a 15 minutes basis. It is
interesting to observe here that the shorter the time intervals on which the \ac{HP} energy demand is to be predicted,
the higher the stochasticity of that prediction. Hence, while shorter bidding intervals in the spot market will allow
for a higher time resolution in the energy production-demand balance, it will also make domestic energy demand
prediction from HVAC systems less accurate.

An interesting point to stress here is that the role of the energy demand prediction in the context of domestic flexible demand based on a heating system is to evaluate the economic cost of running the heating system. If the hourly electricity prices (possibly including VAT and grid fees) are the only economic aspect to take into account, then the expected economic cost of the heating system reads as:
\begin{align}
    \label{eq:SpotCost}
    J = \mathbb E\left[\sum_{k=0}^N \vect \sigma_k \vect P_k\right]
\end{align}
where $\vect P_{0,\ldots,N}$ is the (possibly) stochastic prediction of the average power demand over the hourly price intervals $k$, $\vect \sigma_{0,\ldots,N}$ are the (typically day-ahead) future electricity prices, and the expected value $\mathbb E[.]$ is taken over the sequence $\vect P_{0,\ldots,N}$. In order to estimate correctly a cost in the form \eqref{eq:SpotCost}, it is sufficient to deliver a correct prediction of $\vect P_{0,\ldots,N}$ in terms of its expected values, and readily use it in \eqref{eq:SpotCost} without the expected value operator. Because of its linearity, cost $J$ can also be treated for different devices (heating, ventilation, appliances, etc) in the house separately.

However, the economic cost of domestic energy demand is already in some areas of NordPool are already non-linear, due to peak energy demand penalties. More complex economic costs can be expected in the future, as domestic consumers participate in more complex power markets. An economic cost involving peak energy demand penalties can be put in the generic form:
\begin{align}
    \label{eq:EcoCost}
    J = \mathbb E\left[\Psi\left( \vect P\right) + \sum_{k=0}^N \vect \sigma_k \vect P_k\right]
\end{align}
where $\Psi\left(.\right)$ is a nonlinear, discontinuous function of the sequence $\vect P_{0,\ldots,N}$. For an economic cost function of the form \eqref{eq:EcoCost}, using a simple prediction of the expected demand in \eqref{eq:EcoCost} would not deliver a correct estimation of $J$, and a more detailed prediction of the statistics of the prediction $\vect P_{0,\ldots,N}$ can be beneficial. In addition, while the terms in \eqref{eq:EcoCost} related to the prices $\vect \sigma_{0,\ldots,N}$ can be treated independently for the different devices in the house, the part represented by $\Psi\left( \vect P\right)$ ought to be based on the total average power consumption of the house (i.e. the one measured by the smart meter).

In the following, we will investigate Quantile Regressions (QR), see e.g. \cite{koenkerQuantileRegression2005}, as a way to model the stochasticity of the predictions of $\vect P_{0,\ldots,N}$. QR is a very simple and computationally relatively inexpensive way of building statistical models. QR is based on convex loss functions. If used in conjunction with linearly parametrized models, it yields convex regression problems, which can be solved efficiently to global optimality.

\subsection{Heat-pump power consumption model}
\label{sec:hppower_model}
The house consists of four volumes, each with a different thermal capacity and different thermal losses to the environment. Each volume has a temperature sensor and an indoor heating unit that is connected to two outdoor blocks extracting heat from the air. The heating power in each volume is controlled indirectly via setting the unit ON / OFF labelled $\mathrm{O}_{k,i} \in \{0,1\}$, assigning a target temperature labelled $T_{k,i}^{\text{set}} \in \{16,\ldots,31\}$, and the fan speed $\mathrm{F}_{k,i} \in \{1,2,3,4,5\}$ promoting air mixing in the volume. These settings influence the temperature $T_{k,i}^{\text{volume}} \in \mathbb{R}$ measured in the \ac{HP} volume. We have conducted in-depth investigations on how to best predict the \ac{HP} energy consumption, using methods ranging from simple static linear regression models, auto-regressive models, Hammerstein models, and Deep Neural Network models \cite{belsvik2023machine}. The conclusions of these investigations is that the hourly \ac{HP} can be predicted based on the \ac{HP} setting within the hour, and does not gain much accuracy by using data from previous hours. Hence, the \ac{HP} hourly energy prediction can be treated as a static problem, which does not depend on past data. Furthermore, our observation is that the \ac{HP} settings at a time $k$ can be condensed into a single nonlinear feature, which performs reasonably well to predict the hourly energy consumption when used in a linearly parametrized regression model. The selected nonlinear feature, illustrated in Fig. \ref{fig:HeatPumpFeature} below, reads as
\begin{equation}
    \label{eq:Feature}
    \vect a_{k,i} = \mathrm{O}_{k,i} \cdot \mathrm{F}_{k,i} \cdot \frac{1}{2}\tanh\left(\frac{1}{5}(T_{k,i}^{\text{set}} - T_{k,i}^{\text{volume}}) + 1\right),
\end{equation}
and will also be used in predicting the thermal response of the volumes. Feature \eqref{eq:Feature} consists in making
the \ac{HP} energy demand proportional to the difference between the target temperature $T^{\text{set}}_{k,i}$ and the
measured one $T^{\text{volume}}_{k,i}$, with a smooth saturation term. \textcolor{black}{While other saturation functions
    could be used, the $\tanh$ function is a simple smooth saturation function that is well-behaved in the optimization
    layer of a predictive control scheme. It is cheap to evaluate and we expect alternatives like the softmax function to
    perform worse considering the numerics of the optimization.} The energy demand is weighted by the fan
speed $ \mathrm{F}_{k,i}$, as a higher speed produces a faster dissipation of the heat into the volume, and is activated
or deactivated by the ON / OFF state captured in $\mathrm{O}_{k,i}$. The factor $\frac{1}{5}$ in the argument of the
$\tanh$ function was chosen heuristically, based on our direct observation of the \ac{HP} saturation. We have
investigated the optimization of these parameters for prediction performance, without observing a very large
improvement. \textcolor{black}{This includes nonlinear regression models.}

\begin{figure}[h]
    \center
    \includegraphics[width=1\columnwidth]{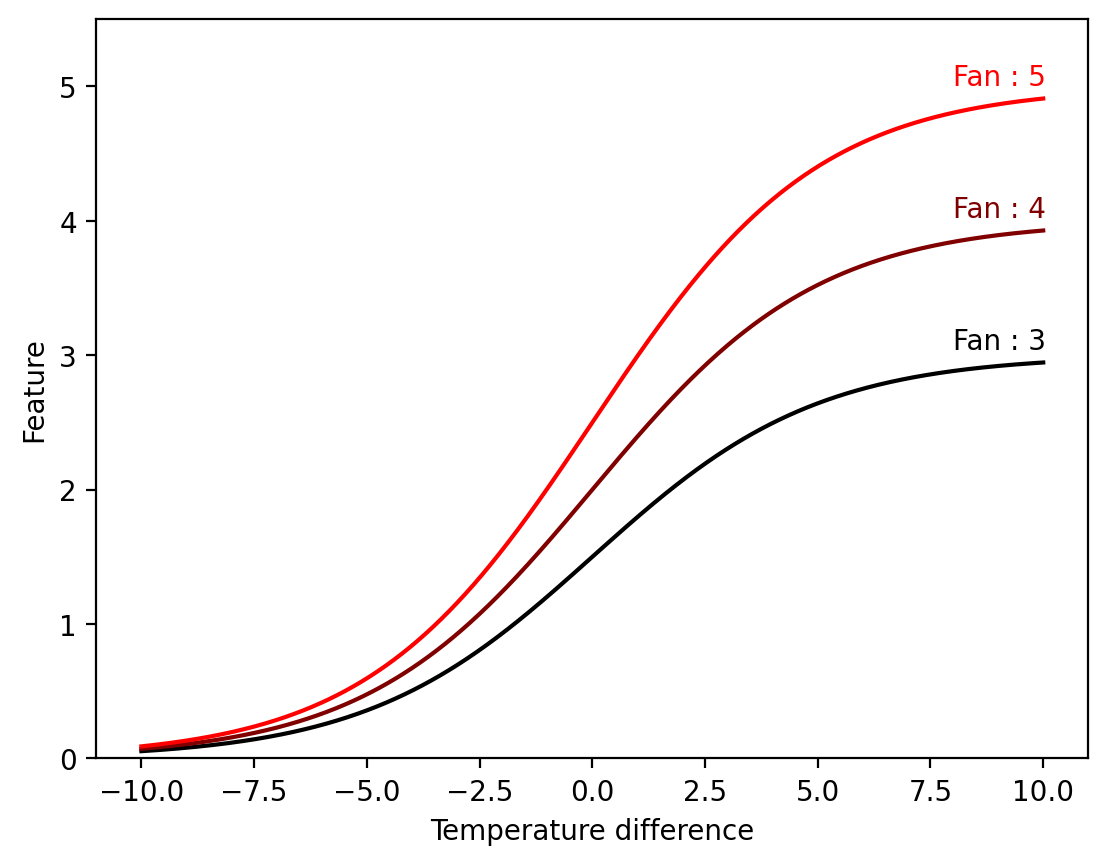}
    \caption{\footnotesize Illustration of the feature function \eqref{eq:Feature} for $\mathrm{O}_{k,i}=1$, modelling the actuation of a heating unit in the experimental house.}
    \label{fig:HeatPumpFeature}
\end{figure}

Using this feature, the heat pumps hourly energy prediction model was linearly parametrized, with:
\begin{align}
    \vect P_h = \alpha_h^\mathrm{out} T^{\text{out}}_h+\sum_{i = 1}^4  \left( \alpha_i^\mathrm{on}   \sum_{k\in H}\mathrm{ON}_{k,i}+\alpha_i^\mathrm{feat} \sum_{k\in H}     \vect a_{k,i} \right)
\end{align}
where $H$ is the set of time indices pertaining to the hourly time interval preceding hour $h$, and $T^{\text{out}}_h$ is the outdoor temperature within that interval, and is treated as constant here.

\begin{figure}
    \center
    \includegraphics[width=1\columnwidth]{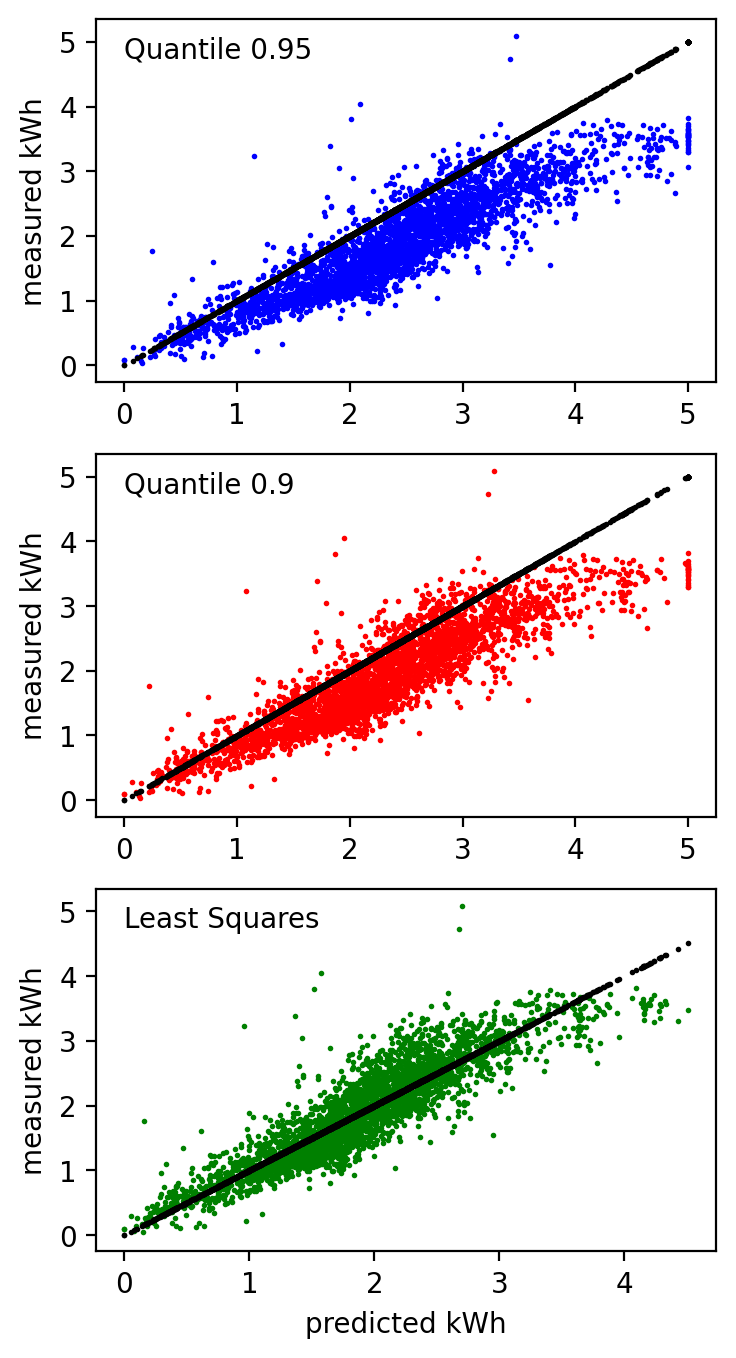}
    \caption{\footnotesize Illustration of the predicted energy demand vs. the measured ones for data in the time interval 2021 Nov. $11^\mathrm{th}$ and 2022 March $1^\mathrm{st}$. The Least-Squares regression (lower graph) delivers a correct expected value, useful for estimating the economic cost of the energy demand due to spot prices. The Quantile regressions with quantiles $0.95$ and $0.9$ deliver models that have 95\%, resp. 90\% empirical probability of upper-bounding the real energy demand. These models are useful to manage economic penalties in a probabilistic way based on exceeding thresholds.}
    \label{fig:PowerRegFit}
\end{figure}

\begin{figure}
    \center
    \includegraphics[width=1\columnwidth]{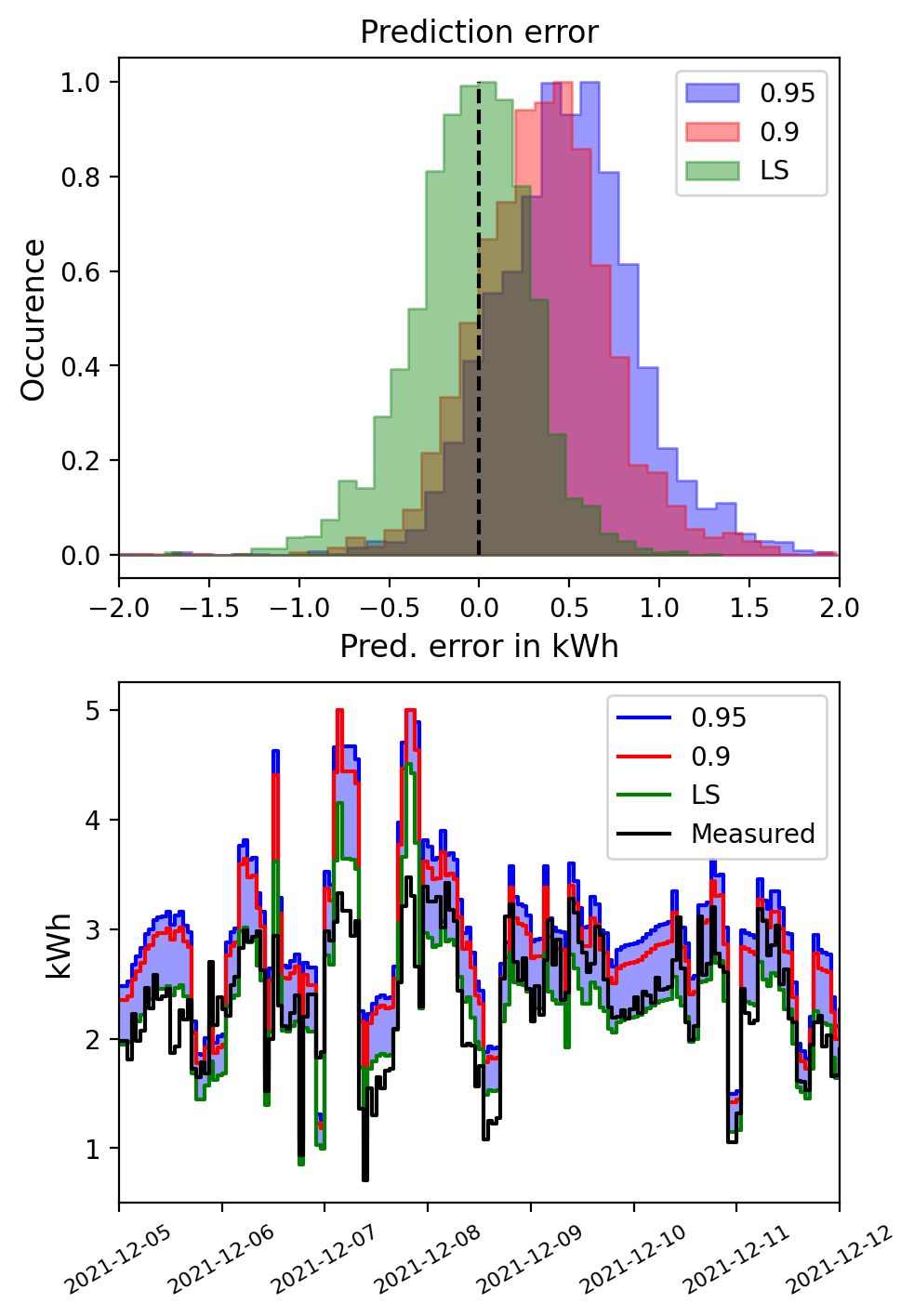}
    \caption{\footnotesize Distribution and time series of the energy demand predictions. The upper graph shows the historgram of the errors of the Least-Squares and Quantile regressions. The lower graph displays a time series of the real energy demand vs. the different models. The Least-Squares model (LS) delivers predictions close to the true energy demand in expected values, but can fail to capture outliers, which can create large economic costs for nonlinear economic penalties. The models obtained from Quantile Regressions offer a viable way to model these outliers.   }
    \label{fig:PowerRegDist}
\end{figure}

The model parameters were identified using first a classic least-squares regression, aiming at delivering a hourly energy prediction model with a correct expected value. Then QR is used to deliver models providing energy demand predictions that have assigned probabilities of being above the true one. The outcome of these regressions are illustrated in Fig. \ref{fig:PowerRegFit}-\ref{fig:PowerRegDist}. Fig. \ref{fig:PowerRegFit} shows the predicted energy demands vs. the measured ones for the three regressions. Fig. \ref{fig:PowerRegDist} upper graph displays the histograms of the three regressions, and the lower graphs displays a time series of the predictions vs. real energy demand.

Because QRs are cheap to compute and based on convex loss functions, they offer a reliable pathway to model the energy demand for economic optmimization. In conjunction with a linearly parametrized regression models, the QR problem is convex and can be solved to global optimality. By forming a family of QR models, one can deliver empirical probabilities that the energy demand will not exceed the different peak energy demand thresholds, which can be used to manage the associated economic cost in a probabilistic but formal manner.



\section{Data-driven predictions of the thermal response}
\label{sec:results}


\begin{figure}
    \center
    \includegraphics[width=1\columnwidth]{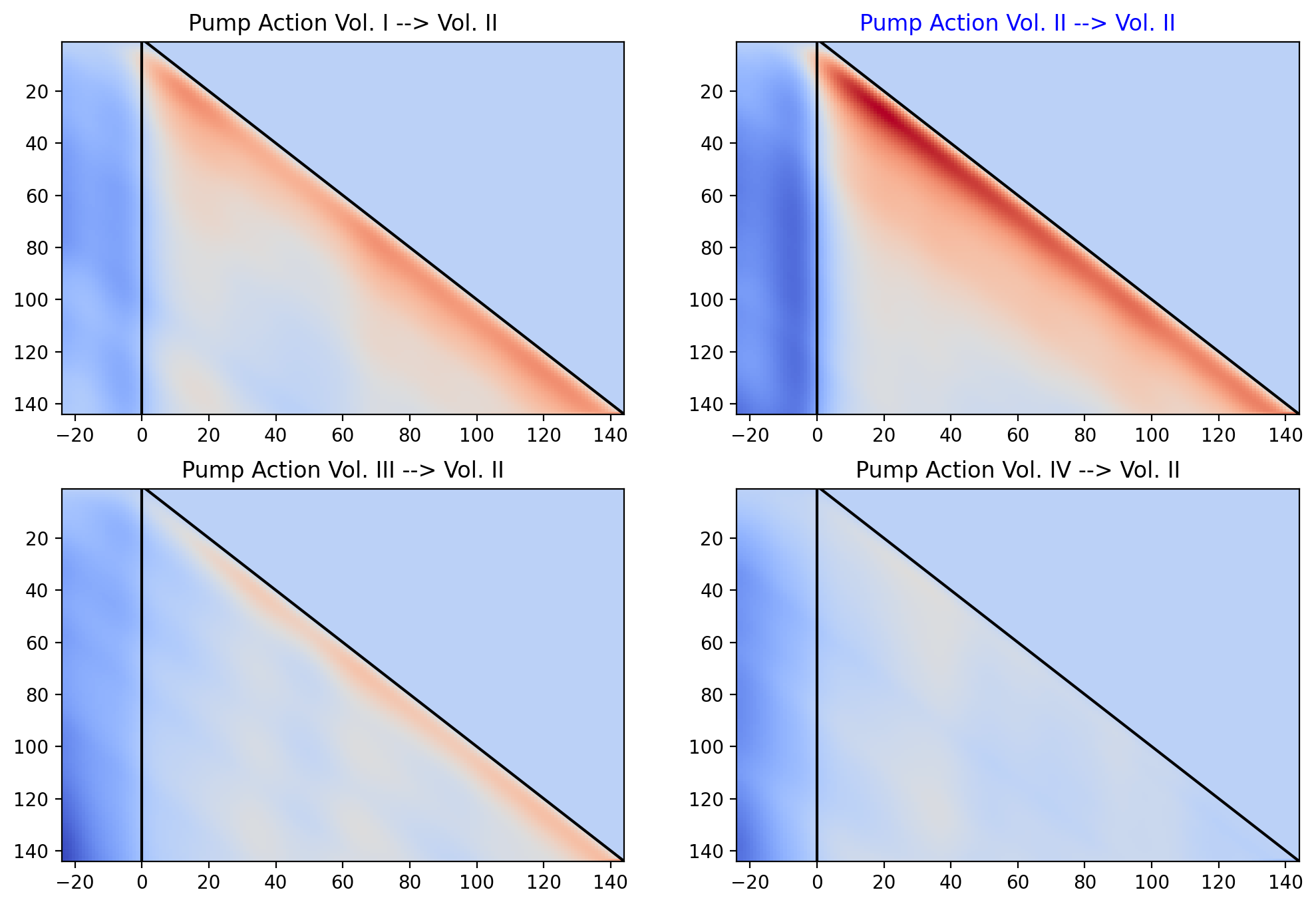}
    \caption{\footnotesize Illustration of the multistep predictor obtained from data in the period 2021 Nov. 15 to 2021 Dec. 15. The heat map displays the effect of the actions \eqref{eq:Feature} of the four heat pumps on the temperature response of volume II. Similar maps are built jointly by the multistep predictor for each volume, and for the influence of external parameters on the volumes.}
    \label{fig:MultiStepFixed:Predictor}
\end{figure}

\begin{figure}[htbp]
    \centering
    \begin{adjustbox}{max width=\columnwidth}
        \includegraphics{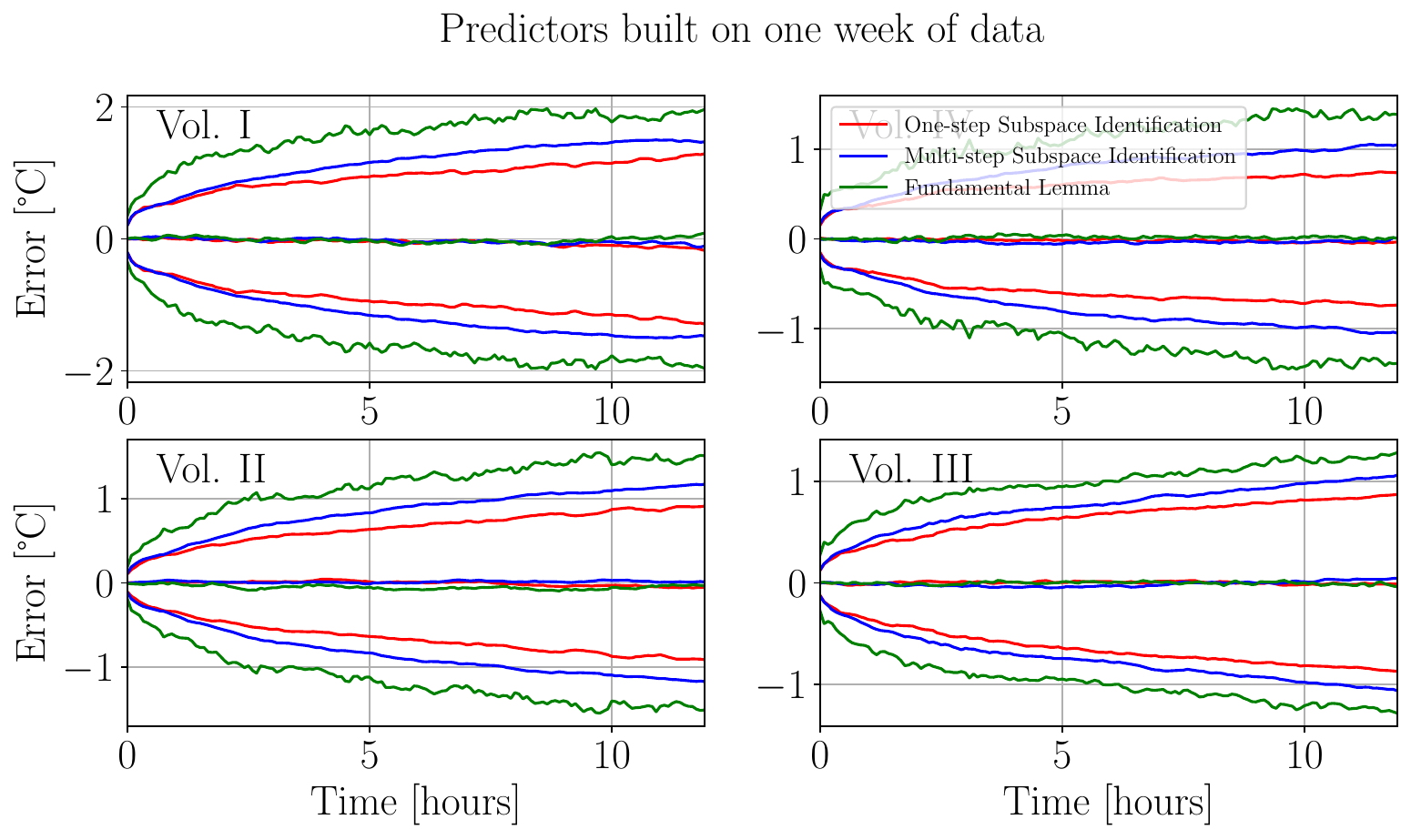}
    \end{adjustbox}
    \begin{adjustbox}{max width=\columnwidth}
        \includegraphics{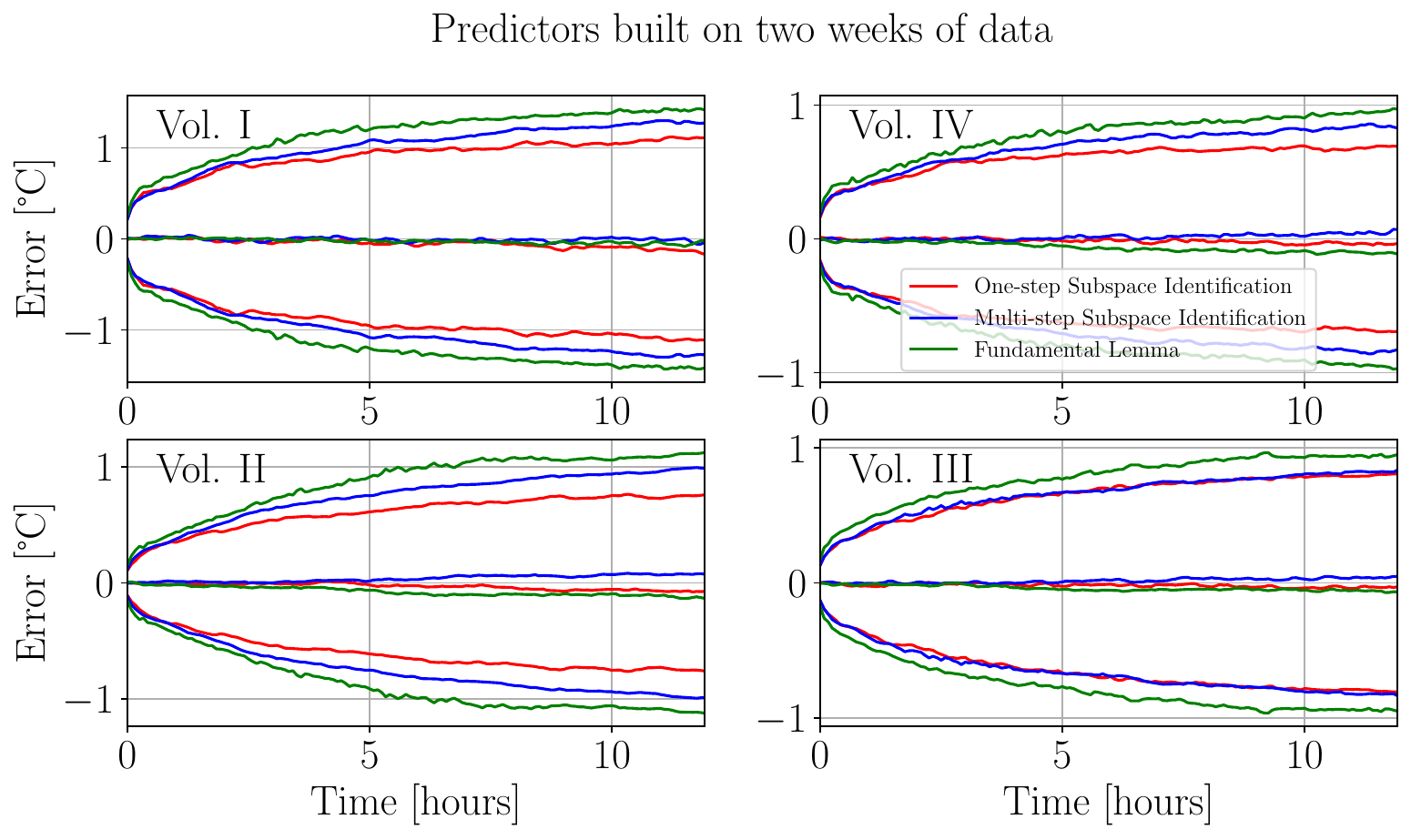}
    \end{adjustbox}
    \caption{Comparison for one week and two weeks of data. In each graph the mid-curve represents the
        mean of the error, and the outer curves represent the standard deviation. All predictors give fairly similar
        results.} \label{fig:compare_one_two}
\end{figure}

\begin{figure}[htbp]
    \centering
    \begin{adjustbox}{max width=\columnwidth}
        \includegraphics{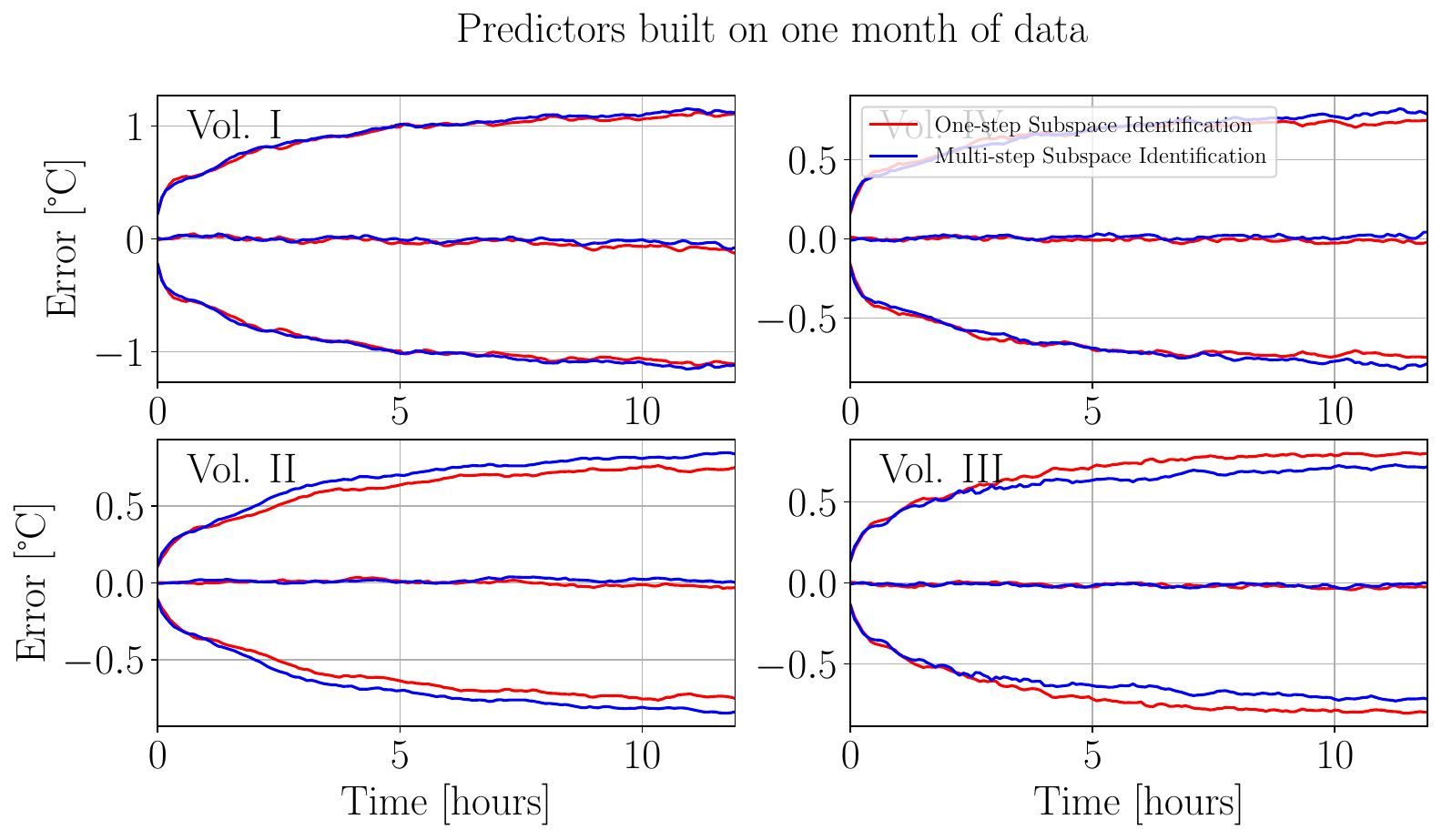}
    \end{adjustbox}
    \caption{Comparison for one month of data. In each graph the mid-curve
        represents the mean of the error, and the outer curves represent the standard deviation.  Building the prediction
        model from one month of data for the Fundamental-Lemma approach was prohibitively expensive, so we only report the
        results for the other methods. Both predictors give fairly similar results.} \label{fig:compare_month}
\end{figure}

In this section, we focus on modeling the thermal dynamics of the house as detailed in \cref{sec:experimental_setup}.
The house is outfitted with heat pumps that the \ac{EMS} manages. This \ac{EMS} interacts with the electrical grid to
receive updates on electricity pricing and demand response signals. It leverages this data to strategically control the
heat pump operations, aiming to minimize electricity expenses while maintaining the thermal comfort of the occupants.
The \ac{EMS} depends on an accurate predictive model of the house thermal and energetic response to the control inputs. These
predictions facilitate the optimized management of the heat pumps within an \ac{EMPC} framework.

As described in \cref{sec:HPPower}, the residence is divided into four distinct volumes, each characterized by its
thermal mass and insulation properties. Each volume is equipped with a temperature sensor and a heating device
linked to the heat pumps. The actuation remains as outlined in \cref{sec:HPPower}, wherein the heat pump
output at time step $k$ is indirectly manipulated by setting the $i$th unit to $O_{k,i}$ (ON/OFF), setting a desired
temperature $T^\text{set}_{k,i}$, and adjusting the fan speed $F_{k,i}$ to facilitate air circulation within the zone.
These control variables form the features $\mathbf{a}_{k,i}$, with $i\in\{1,2,3,4\}$, which we employ as the inputs to the
predictive models for the house thermal response, as defined in \cref{eq:Feature}. Additionally, the external
temperature $T^{\text{out}}_k \in \mathbb{R}$ is considered an exogenous input. The input vector $\mathbf{u}_k \in
    \mathbb{R}^5$ and output vector $\mathbf{y}_k \in \mathbb{R}^4$ are then constructed  as
\begin{equation}
    \mathbf{u}_k = \left[\mathbf{a}_{k,i}\quad T_k^{\text{out}}\right]^\top, \quad \mathbf{y}_k = \left[T_{k,i}^\text{volume}\right]^\top, \quad i \in {1,2,3,4},
\end{equation}
where $\mathbf{y}_k$ collects the temperatures of each volume.

We employ the data-driven methodologies outlined in Section \ref{sec:theoretical_background} to construct predictive
models for the thermal response of the house. Our evaluation of these models focuses on comparing their predictive
accuracy and computational efficiency, as well as their practical use in predictive control. To develop the models, we
utilize datasets of varying sizes, specifically from durations of one week, two weeks, and one month. Using the
resulting models, we predict the thermal behavior of the house for a forthcoming 12-hour period directly following the data sets, basing the predictions
on the initial conditions derived from data of the two hours immediately prior (these two hours are not part of the
model-building dataset). By iterating this process at three-hour intervals across the complete dataset, we obtain a
large number of predictions that we use to evaluate each method based on the statistical mean and variance of the resulting
prediction error. This enables us to assess the models under diverse weather scenarios and varying occupancy patterns.

\subsection{Data size and computational complexity}
Indoor temperature readings, denoted by $T_{k,i}^{\text{volume}}$, and the inputs to construct $ \mathbf{a}_{k,i}$ are
sampled at a 5-minute interval. Conversely, outdoor temperatures $ T^{\text{out}}_k$ are recorded hourly and are
subsequently linearly interpolated to match the 5-minute sampling. Despite the presence of noise in the
dataset, we did not apply any filtering, assuming that the high volume of data compensates for any noise-induced
anomalies. Shortfalls in data continuity, if under a 10-minute threshold, are rectified via linear interpolation; gaps
surpassing this duration result in data exclusion. This preparation yields a dataset conducive to approximately 1,100
modeling and thermal response prediction iterations. Taking into account the 5-minute sampling rate and two hours for
initial conditions and twelve hours for prediction, we employ $ N_p = 24$ and $ N_f = 144$ as dimensions for the data
across all prediction methods. The assembled datasets for one week, two weeks, and one month correspond to values of $
    \bar{N}$ at $ \{2016, 4032, 8064\}$ respectively.

The time required for a single cycle of model building and prediction on a conventional office laptop varies with the
computational complexity of the method: the one-step predictor completes within seconds, the multi-step predictor
requires about 10-20 seconds depending on the size of the data window, and the Fundamental-Lemma based predictor extends to about four minutes for a data window of two weeks, and is difficult to complete for a one month window. This notable
discrepancy in processing time can be attributed to the reliance of the Fundamental-Lemma predictor on solving a dense \ac{QP}
at each sampling time of the \ac{EMS} controller. We obtain the solutions via the Python-based \texttt{cvxpy} package~\cite{diamond2016cvxpy}, which is not
specifically developed for rapid computation. In contrast, the other methods benefit from direct least-squares
approaches, solvable through standard linear algebra libraries. 
In general, the one-step and multi-step predictions are computationally efficient, while the
Fundamental-Lemma predictor yields a significantly higher computational complexity, irrespective of the employed \ac{QP}
solver. One consequence is that the one-step and multi-step predictions can be solved on arbitrarily large data of windows, while the Fundamental-Lemma predictor cannot due to its increasing computational complexity over the window size.

More importantly, the one-step / multi-step predictors can be assembled infrequently and independently of the control loop (e.g. once a day or less) and used for Predictive Control for a certain period of time. In contrast, the LF approach has to solve the regression at every sampling time of the \ac{EMS} control system, and cannot do it independently of the control.

Our implementation of the Fundamental-Lemma predictor incorporates a composite \(\ell_1/\ell_2\) regularization
function, defined as:
\begin{equation}
    \Pi(\mathbf{g}) = \lambda_1 \|\mathbf{g}\|_1 + \lambda_2 \|\mathbf{g}\|_2^2,
\end{equation}
with $\lambda_1$ and $\lambda_2$ were set to 100 and 1, respectively. The selection of these parameters leverages the
$\ell_1$ norm propensity for inducing sparsity, which is well-regarded in the realm of data-driven predictive
control~\cite{Dorfler2022}. In contrast, an overly dominant $\ell_2$ norm would lead to a predictor similar to the
multi-step variant obtainable through a least-squares approach as outlined in~\cref{eq:phi_N_f-step}. However, as long
as $\ell_1$ regularization terms are in the formulation, the solution at the cost of solving a dense \ac{QP} instead of
computing it via a linear least-squares problem.

Moreover, the Fundamental Lemma predictor in the discussed form in general does not exploit the block-Toeplitz
structure, see \cref{fig:MultiStepFixed:Predictor} and thus always yields non-causal predictions, hence ignoring a
fundamental prior knowledge readily available for the system.

\subsection{Prediction accuracy}

We evaluate the prediction accuracy of the three methods by comparing the predicted temperatures to the actual
measurements over all iterations for predictors based one week, two weeks and one month of data. The Fundamental Lemma
approach was not done on a month of data, due to its computational cost for that window size. A comparison of the three
methods based on the mean and standard deviation is shown in \cref{fig:compare_one_two} and \cref{fig:compare_month}.

The prediction accuracy of all predictors improves for increasing size of the dataset, though the performance gain appears to stop increasing beyond one month of data (not displayed here for the sake of brevity).  The increase in prediction error
from small magnitudes in the beginning of the prediction horizon to larger magnitudes at the end of the prediction appears
similar among all methods and datasets. Even for one week, the bias seems to be small, indicated by a mean prediction error
close to zero among the prediction horizon. The standard deviation of the prediction errors is mostly below $1\si{\celsius}$,
except for Vol I, where the air circulation is more stochastic due to its large size. The one-step predictor and multi-step predictor have similar prediction accuracy, while the
Fundamental-Lemma predictor has a slightly higher prediction error and generally more noisy predictions which may be
attributed to the significant $\ell_1$ regularization.

The improvement from using two weeks of data to one month is not as significant as one might expect, and one week of data
appears to be sufficient to obtain fairly accurate predictions. This suggests that the one-step or multi-step predictors are applicable on a short data size, which is important for domestic flexible demand, where the \ac{EMS} system ought to be operational shortly after its installation.

We finally ought to stress that while the one-step and multi-step predictors have similar performance, the multi-step predictor offers significant advantages in allowing one to build more advance models capturing the statistics of the temperature predictions, e.g. using the Quantile Regression techniques shown in Sec.~\ref{sec:HPPower}.

\subsection{Predictive control}

The aim of the predictive models is to facilitate the control of heat pumps within an \ac{EMPC} framework. An in-depth
discussion of the resulting predictive control schemes is outside the scope of this chapter. However, we briefly discuss
the practical implications of utilizing the different prediction methods for optimization. An \ac{EMPC} scheme is based on a receding horizon
optimization problem that is solved at each time step. Generally, it is favorable to have a long prediction horizon to
capture the long-term effects of control inputs. However, the computational complexity of the optimization problem
increases with the length of the prediction horizon. Thus, it is beneficial to use predictors with a small computational
footprint. The one-step and multi-step predictors from subspace identification are well-suited for this purpose. The
one-step predictor is computationally efficient, but is more prone to prediction biases (albeit this is not apparent in our experience). In addition, the multi-step predictor allows
for the building of statistics over the prediction horizon, thus allowing for describing the uncertainty of the prediction beyond the expected value. This
enables the \ac{EMPC} scheme to account for the stochasticity of the prediction.

The predictor based on the Fundamental Lemma exhibits a much higher computational complexity. Its lower prediction
accuracy compared to the other methods renders it less apt for predictive control for flexible demand. Moreover, there is
an ongoing debate within the community regarding the efficacy of the Fundamental Lemma for predictive control.  A key
distinction between the Fundamental Lemma and the subspace identification approaches is their treatment of model
construction and the control problem, referred to as direct and indirect system identification~\cite{Dorfler2022}. The
Fundamental Lemma merges these into a single optimization problem (direct), while the subspace identification maintains a clear
delineation between the two (indirect). A crucial implication of this difference is that the Fundamental Lemma necessitates a
reiteration of the model construction with each control update. In contrast, the subspace identification requires a
one-time model build, subsequently utilized for control updates. The model built can be infrequently updated, outside the control loop. Hence the interval for reconstructing the model in subspace
identification is a variable design parameter, whereas it is inherently prescribed in the Fundamental Lemma approach.
This inflexibility is a notable disadvantage for the Fundamental Lemma, especially given the computational demands of
constructing the model.

It is moreover argued that balancing the regularization term with the performance criterion of the control problem in the Fundamental Lemma approach is
challenging and that the weighting between the two objectives needs to be tuned extensively \cite{Natale2022LessonsLF}, making a reliable large-scale application of the approach very doubtful.
The predictive accuracy is guaranteed to decrease when the regularization term of the predictor is traded off against
the performance criterion of the control problem. This is a significant drawback for the Fundamental-Lemma approach, as
the predictive accuracy is crucial for the performance of the \ac{EMPC} scheme. The subspace identification approach do
not suffer from this drawback, as the predictive accuracy is not affected by the control problem.


\section{Conclusion}
\label{sec:conclusion}

In this chapter, we have addressed the demand response problem for a typical norwegian household amid the shift of the power system towards
a renewable energy. We introduced a data-driven methodology to forecast the house thermal dynamics in
response to control actions, using auto-regressive methods, methods from subspace identification and the Fundamental Lemma. Through comparative
analysis, we evaluated the predictive accuracy and computational demands of these techniques. Our findings indicate that
the auto-regressive and subspace identification method are computationally more efficient than Fundamental Lemma approach, and also deliver more accurate
predictions. Further, we discussed the implications of these methodologies in the context of predictive control,
deducing that predictors founded on subspace identification are more appropriate for such applications.

\begin{acknowledgement}
    Dirk Reinhardt, Wenqi Cai, and Sebastien Gros are supported by the Research Council of Norway through the project ``SARLEM: Safe Reinforcement Learning using MPC'' (grant number 300172).
\end{acknowledgement}


\bibliographystyle{spmpsci}
\bibliography{main}
\end{document}

%% file: acronyms.tex
\acrodef{SI}{subspace identification}
\acrodef{ARX}{autogressive models with exogenous inputs}
\acrodef{LTI}{linear time-invariant}
\acrodef{FL}{fundamental lemma}
\acrodef{SPC}{subspace predictive control}
\acrodef{OSSI}{one-step subspace identification}
\acrodef{COP}{coefficient of performance}
\acrodef{IoT}{internet of things}
\acrodef{HAN}{Home Area Network}
\acrodef{IR}{infrared}
\acrodef{TVOC}{total volatile organic compounds}
\acrodef{PVGIS}{Photovoltaic Geographical Information System}
\acrodef{API}{application programming interface}
\acrodef{HVDC}{high-voltage direct current}
\acrodef{HVAC}{heating, ventilation, and air conditioning}
\acrodef{EV}{electric vehicle}
\acrodef{EMPC}{Economic Model Predictive Control}
\acrodef{DRL}{Deep Reinforcement Learning}
\acrodef{TES}{thermal energy storage}
\acrodef{HP}{heat pump}
\acrodef{EMS}{energy management system}
\acrodef{QP}{quadratic program}
\acrodef{MES}{Multi-Energy Systems}
\acrodef{ESI}{Energy System Integration}

%% file: figures/spot.tex
\begin{tikzpicture}
    \begin{axis}[
            ybar,
            /pgf/number format/.cd,
            use comma,
            1000 sep={},
            xlabel={Øre/kWh},
            ylabel={Hours},
            x label style={at={(axis description cs:0.5,-0.1)},anchor=north},
            xtick distance=50,
            ytick distance=100,
            width=\columnwidth, 
            height=6cm,
            axis y line*=left,
            axis x line*=bottom,
            ymin=0,
            xmin=0,
            grid=both,
        ]

        \addplot +[red,
        fill opacity=0.5,
        hist={bins=100,
                data min=0,
                data max=220,
            }
        ] table[x=Price, col sep=comma] {figures/spotmarket_Trheim_no_time.csv};
        \addlegendentry{Trondheim}

        \addplot +[black,
        fill opacity=0.5,
        hist={bins=100,
                data min=0,
                data max=300,
            }
        ] table[x=Price, col sep=comma] {figures/spotmarket_Oslo_no_time.csv};
        \addlegendentry{Oslo}
    \end{axis}
\end{tikzpicture}
\begin{tikzpicture}
    \begin{axis}[
            axis y line*=left,
            axis x line*=bottom,
            date coordinates in=x,
            xmin=2021-01-01 00:00,
            xmax=2022-02-01 00:00,
            xtick distance=31,
            ymin=0,
            grid=both,
            xticklabel style={
                    rotate=90,
                    anchor=near xticklabel,
                },
            xticklabel=\month.\year,
        ]
        \addplot[semithick, red] table [col sep=comma,x=Time_start,y=Price] {figures/spotmarket_Trheim.csv};
        \addplot[semithick, black] table [col sep=comma,x=Time_start,y=Price] {figures/spotmarket_Oslo.csv};
    \end{axis}
\end{tikzpicture}

%% file: figures/outdoor_temperature.tex
\begin{tikzpicture}
    \begin{axis}[
            ybar,
            /pgf/number format/.cd,
            use comma,
            1000 sep={},
            xlabel={Temperature [$^\circ$C]},
            ylabel={Number of Measurements [-]},
            x label style={at={(axis description cs:0.5,-0.1)},anchor=north},
            xtick distance=10,
            ytick distance=100,
            width=\columnwidth, 
            height=6cm,
            axis y line*=left,
            axis x line*=bottom,
            ymin=0,
            xmin=-25,
            xmax=+25,
            grid=both,
        ]

        \addplot +[blue,
        fill opacity=0.3,
        hist={bins=100,
                data min=-30,
                data max=+30,
            }
        ] table[x=Temperature, col sep=comma] {figures/outdoor_temperature.csv};
        \addlegendentry{Other}
    \end{axis}
\end{tikzpicture}

%% file: figures/power_consumption.tex
\begin{tikzpicture}
    \begin{axis}[
            ybar,
            /pgf/number format/.cd,
            use comma,
            1000 sep={},
            xlabel={Power [kW]},
            ylabel={Number of Measurements [-]},
            x label style={at={(axis description cs:0.5,-0.1)},anchor=north},
            xtick distance=1,
            ytick distance=500,
            width=\columnwidth, 
            height=6cm,
            axis y line*=left,
            axis x line*=bottom,
            ymin=0,
            xmin=0,
            xmax=4,
            grid=both,
        ]

        \addplot +[blue,
        fill opacity=0.3,
        hist={bins=100,
                data min=0,
                data max=5,
            }
        ] table[x=tibber_up_consumption, col sep=comma] {figures/tibber_up_consumption.csv};
        \addlegendentry{Other}
        \addplot +[red,
        fill opacity=0.3,
        hist={bins=100,
                data min=0,
                data max=5,
            }
        ] table[x=tibber_pumps_consumption, col sep=comma] {figures/tibber_pumps_consumption.csv};
        \addlegendentry{Heat pumps}
    \end{axis}
\end{tikzpicture}